\documentclass[11pt]{article}
\usepackage{amssymb}
\usepackage{amsmath}
\usepackage{amsfonts}
 \usepackage{pifont}
\usepackage{amsmath,amsfonts,amssymb,theorem,graphicx,listings,txfonts}
\usepackage{graphics}
\usepackage{flafter}
 \usepackage{indentfirst}
\usepackage{epsfig}
  \usepackage{mathrsfs}
 \usepackage{cite}
 \usepackage[algosection,linesnumbered,ruled,lined]{algorithm2e}
 \usepackage[linesnumbered,ruled,lined]{algorithm2e}
\newtheorem{theorem}{Theorem}[section]

\newtheorem{definition}[theorem]{Definition}
\newtheorem{example}[theorem]{Example}

\renewcommand{\arraystretch}{1}

\title{Knowledge reduction of dynamic covering decision information systems with varying attribute values}

\author
{Mingjie Cai$^{a}$ \\
\small {$^{a}$ College of Mathematics and Econometrics, Hunan University}\\
\small {Changsha, Hunan 410082, P.R. China}\\
}
\date{}

\begin{document}
\maketitle \baselineskip=17pt
\begin{center}
\begin{quote}
{{\bf Abstract.}
Knowledge reduction of dynamic covering information systems involves with the time in practical situations. In this paper, we provide incremental approaches to computing the type-1 and type-2 characteristic matrices of dynamic coverings because of varying attribute values. Then we present incremental algorithms of constructing the second and sixth approximations of sets by using characteristic matrices. We employ experimental results
to illustrate that the incremental approaches are effective to calculate
approximations of sets in dynamic covering information systems. Finally, we perform knowledge reduction of dynamic covering information systems with the incremental approaches.

\\}
\end{quote}
\end{center}
\renewcommand{\thesection}{\arabic{section}}

\section{Introduction}

Covering approximation spaces, as generalizations of classical approximation spaces based on equivalence relations, have attracted more attentions, and a great deal of approximation operators have been proposed for knowledge reduction of covering approximation spaces. Nowadays, covering-based rough set theory\cite{Tan1,Tan2,Wang5, Wang7,Yang,Yang3,Zhang2,Zhu1,Zhu4,Zhu5,Zakowski1} are being enriched with the development of computer sciences and related theories.

To our best knowledge, there exist many lower and upper approximation operators for covering approximation spaces, and their basic properties are investigated concretely by researchers. Especially, Wang et al.\cite{Wang1} studied the second and sixth lower and upper approximation operators of covering approximation spaces and proposed effective approaches to computing the second and sixth lower and upper approximations of sets by using characteristic matrices. In practice, dynamic covering approximation spaces are variations of the time. For example, two specialists A and B  decided the quality of five cars $U=\{A, B,C,D,E\}$ as follows: $good=\{A,C\},middle=\{C,E\},bad=\{B,D,E\}$, and $(U,\mathscr{C})$ is a covering approximation space, where $\mathscr{C}=\{good,middle,bad\}$. With the variation of the time, the specialists find
that the quality of $C$ is very bad, and $(U,\mathscr{C})$ is revised into dynamic covering approximation space $(U,\mathscr{C}^{\ast})$, where $\mathscr{C}^{\ast}=\{good^{\ast},middle^{\ast},bad^{\ast}\}$, $good^{\ast}=\{A\},middle^{\ast}=\{E\}$ and $bad^{\ast}=\{B,C,D,E\}$.
Accordingly, the characteristic matrice of $\mathscr{C}$ changes into that of
$\mathscr{C}^{\ast}$. Since it is time-consuming to compute the characteristic matrice in large-scale covering approximation space, it costs more time to construct the characteristic matrice of large-scale dynamic coverings for computing approximations of sets. Until now, Lang et al.\cite{Lang1, Lang2} presented incremental approaches to computing approximations of sets in dynamic covering approximation spaces, in which object sets are variations of the time. But little attention has been paid to dynamic covering approximation spaces, in which elements of coverings are variations of the time. Therefore, it is of interest to study how to compute approximations of sets in dynamic covering approximation spaces when varying attribute values.

Many researchers\cite{Chen,Li,Liu,Shu,
Wang,Yang,Yang2,Chen1,Chen3,Li1,Li11,Li12,Liang1,
Liu1,Luo1,Luo2,Luo3,
Liu2,Wang3,Zhang1,Zhang4,Zhang,Shan1,Shu1} have investigated knowledge reduction of dynamic information systems with incremental approaches.
For example, when coarsening and refining attribute values and varying attribute sets, Chen et al.\cite{Chen1,Chen3,Chen} constructed approximations of sets which provides an effective approach to knowledge reduction of dynamic information systems.
Based on characteristic relations, Li, Ruan and Song\cite{Li11} extended rough sets for incrementally updating decision rules which handles dynamic maintenance of decision rules in data mining.
Liu et al.\cite{Liu1,Liu2,Liu} presented incremental approaches for knowledge reduction of dynamic information systems and dynamic incomplete information systems.
From the perspective of knowledge engineering and neighborhood systems-based rough sets, Yang, Zhang, Dou and Yang\cite{Yang2} studied the neighborhood system for knowledge reduction of incomplete information systems.
Zhang, Li and Chen\cite{Zhang} presented matrice-based approaches for computing the approximations, positive, boundary and negative regions in composite information systems.
Illustrated by existing researches, the incremental approaches are effective to conduct knowledge reduction of dynamic information systems, which reduces the computation times greatly. It motivates us to compute approximations of sets in dynamic covering approximation spaces and knowledge reduction of dynamic covering information systems by using incremental approaches.

The purpose of this paper is to further study knowledge reduction of dynamic covering information systems when
varying attribute values. First, we investigate structures of the
characteristic matrices of dynamic covering approximation spaces when varying attribute values and present  incremental approaches to
computing characteristic matrices of dynamic coverings. We employ several examples to illustrate that the
process of calculating the characteristic matrices is simplified
greatly by utilizing the incremental approaches. Second, we provide
incremental algorithms for constructing the characteristic
matrices-based approximations of sets in dynamic covering approximation
spaces when varying attribute values. We also compare the time complexities of the incremental algorithms with those of non-incremental
algorithms. Third, we perform experiments on
ten dynamic covering approximation spaces generated randomly. The
experimental results illustrate that the proposed approached are
effective to calculate approximations of sets with respect to the
variation of attribute values. We also employ examples to show that how to conduct knowledge reduction of dynamic covering information systems with the incremental approaches.

The rest of this paper is organized as follows: Section 2 briefly
reviews the basic concepts of covering-based rough set theory. In
Section 3, we introduce incremental approaches to computing the
characteristic matrices of dynamic coverings when varying attribute values. Section 4 presents non-incremental and incremental algorithms of
calculating the second and fifth lower and upper approximations of
sets by using the characteristic matrices. Section 5 performs
experiments to show that the incremental approaches are effective to
compute approximations of sets in dynamic covering approximation
spaces. Section 6 is devoted to knowledge reduction of dynamic covering information systems with the incremental approaches. We conclude the paper in Section 7.

\section{Preliminaries}

A brief summary of related concepts in covering-based rough sets is given in this section.

Let $U$ be a finite universe of discourse, and $\mathscr{C}$ is a
family of subsets of $U$. If none of elements of $\mathscr{C}$ is
empty and $\bigcup\{C|C\in \mathscr{C}\}=U$, then $\mathscr{C}$ is
referred to as a covering of $U$. In addition, $(U,\mathscr{C})$ is
called a covering approximation space if $\mathscr{C}$ is a covering
of $U$.

\begin{definition}\cite{Wang1}
Let $U=\{x_{1},x_{2},...,x_{n}\}$ be a finite universe, and
$\mathscr{C}=\{C_{1},C_{2},...,C_{m}\}$ a covering of $U$. For any
$X\subseteq U$, the second, fifth and sixth upper and lower approximations
of $X$ with respect to $\mathscr{C}$, respectively, are defined as
follows:

$(1)$ The second upper and lower approximations of $X$:
\begin{equation*}
SH_{\mathscr{C}}(X)=\bigcup\{C\in\mathscr{C}\mid C\cap X\neq
\emptyset\},\quad SL_{\mathscr{C}}(X)=[SH_{\mathscr{C}}(X^{c})]^{c};
\end{equation*}

$(2)$ The fifth upper and lower approximations of $X$:
\begin{equation*}
IH_{\mathscr{C}}(X)=\{x\in U\mid N(x)\cap X\neq \emptyset\},\quad
IL_{\mathscr{C}}(X)=\{x\in U\mid N(x)\subseteq X\};
\end{equation*}

$(3)$ The sixth upper and lower approximations of $X$:
\begin{equation*}
XH_{\mathscr{C}}(X)=\bigcup\{N(x)\mid N(x)\cap X\neq
\emptyset\},\quad XL_{\mathscr{C}}(X)=\bigcup\{N(x)\mid
N(x)\subseteq X\}.
\end{equation*}

\end{definition}
\begin{definition}\cite{Wang1}
Let $\mathscr{C}$=$\{C_1,...,C_m\}$ be a family of subsets of a finite set $U$=$\{x_1,...,x_n\}$.
We define $M_\mathscr{C}=(a_{ij})_{n\times m}$, where
$a_{ij}=\left\{
\begin{array}{ccc}
1,&{\rm}& x_{i}\in C_{j},\\
0,&{\rm}& x_{i}\notin C_{j}.
\end{array}
\right. $
\end{definition}

\begin{definition}\cite{Wang1}
Let $U=\{x_1,...,x_n\}$, $A\subseteq U$. We define the
characteristic function as $\mathcal {X}_{A}
=\left[\begin{array}{cccccc}
  a_{1}&a_{2}&.&.&. & a_{n} \\
  \end{array}
\right]^{T}$, where $a_{i}=\left\{
\begin{array}{ccc}
1,&{\rm}& x_{i}\in A,\\
0,&{\rm}& x_{i}\notin A.
\end{array}
    i=1,\cdots,n.
\right. $
\end{definition}

\begin{definition}\cite{Wang1}
Let $U=\{x_{1},x_{2},...,x_{n}\}$ be a finite universe,
$\mathscr{C}=\{C_{1}, C_{2}, ..., C_{m}\}$ a covering of $U$, and
$M_{\mathscr{C}}=(a_{ij})_{n\times m}$ the matrice representation of
$\mathscr{C}$. Then $\Gamma(\mathscr{C})=M_{\mathscr{C}}\cdot
M_{\mathscr{C}}^{T}=(b_{ij})_{n\times n}$ is called the type-1
characteristic matrix of $\mathscr{C}$, where $a_{ij}=\left\{
\begin{array}{ccc}
1,&{\rm}& x_{i}\in C_{j};\\
0,&{\rm}& x_{i}\notin C_{j}.\end{array} \right.$ and
$b_{ij}=\bigvee^{m}_{k=1}(a_{ik}\cdot a_{jk})$.
\end{definition}

\begin{definition}\cite{Wang1}
Let $U=\{x_{1},x_{2},...,x_{n}\}$ be a finite universe,
$\mathscr{C}=\{C_{1}, C_{2}, ..., C_{m}\}$ a covering of $U$, and
$M_{\mathscr{C}}=(a_{ij})_{n\times m}$ the matrice representation of
$\mathscr{C}$. Then $\prod(\mathscr{C})=M_{\mathscr{C}}\odot
M_{\mathscr{C}}^{T}=(c_{ij})_{n\times n}$ is called the type-2 characteristic matrice of
$\mathscr{C}$, where $c_{ij}=\bigwedge^{m}_{k=1}(a_{jk}-a_{ik}+1).$
\end{definition}

By Definitions 2.4 and 2.5, the second and fifth lower and upper
approximation operators are axiomatized equivalently as follows.

\begin{definition}\cite{Wang1}\label{def:operator}
Let $U=\{x_{1},x_{2},...,x_{n}\}$ be a finite universe,
$\mathscr{C}=\{C_{1},C_{2},...,C_{m}\}$ a covering of $U$, and
$\mathcal {X}_{X}$ the characteristic function of $X$ in $U$. Then
$${X}_{SH(X)}=\Gamma(\mathscr{C})\cdot \mathcal {X}_{X},
\mathcal {X}_{SL(X)}=\Gamma(\mathscr{C})\odot \mathcal {X}_{X};
\mathcal {X}_{IH(X)}=\prod(\mathscr{C})\cdot \mathcal {X}_{X},\quad
\mathcal {X}_{IL(X)}=\prod(\mathscr{C})\odot \mathcal {X}_{X}.
$$
\end{definition}

\begin{definition}\cite{Lang1}
Let $(U,\mathscr{D}\cup U/d)$ be a covering decision information
system, where $\mathscr{D}=\{\mathscr{C}_{i}|i\in I\}$,
$U/d=\{D_{i}|i\in J\}$, I and J are indexed sets. We define
$\mathscr{P}\subseteq \mathscr{D}$ as the type-1 reduct of
$(U,\mathscr{D}\cup U/d)$ if it satisfies

$(1)$ $\Gamma(\mathscr{D})\cdot \mathcal
{X}_{D_{i}}=\Gamma(\mathscr{P})\cdot \mathcal {X}_{D_{i}},
\Gamma(\mathscr{D})\odot \mathcal
{X}_{D_{i}}=\Gamma(\mathscr{P})\odot \mathcal {X}_{D_{i}}, \forall
i\in J;$

$(2)$ $\Gamma(\mathscr{D})\cdot \mathcal
{X}_{D_{i}}\neq\Gamma(\mathscr{P^{'}})\cdot \mathcal {X}_{D_{i}},
\Gamma(\mathscr{D})\odot \mathcal
{X}_{D_{i}}\neq\Gamma(\mathscr{P^{'}})\odot \mathcal {X}_{D_{i}},
\forall \mathscr{P^{'}}\subset \mathscr{P}.$
\end{definition}

\begin{definition}\cite{Lang1}
Let $(U,\mathscr{D}\cup U/d)$ be a covering decision information
system, where $\mathscr{D}=\{\mathscr{C}_{i}|i\in I\}$,
$U/d=\{D_{i}|i\in J\}$, I and J are indexed sets. We define
$\mathscr{P}\subseteq \mathscr{D}$ as the type-2 reduct of
$(U,\mathscr{D}\cup U/d)$ if it satisfies

$(1)$ $\prod(\mathscr{D})\cdot \mathcal
{X}_{D_{i}}=\prod(\mathscr{P})\cdot \mathcal {X}_{D_{i}},
\prod(\mathscr{D})\odot \mathcal {X}_{D_{i}}=\prod(\mathscr{P})\odot
\mathcal {X}_{D_{i}}, \forall i\in J;$

$(2)$ $\prod(\mathscr{D})\cdot \mathcal
{X}_{D_{i}}\neq\prod(\mathscr{P^{'}})\cdot \mathcal {X}_{D_{i}},
\prod(\mathscr{D})\odot \mathcal
{X}_{D_{i}}\neq\prod(\mathscr{P^{'}})\odot \mathcal {X}_{D_{i}},
\forall \mathscr{P^{'}}\subset \mathscr{P}.$
\end{definition}

\begin{definition}\cite{Wang1}
Let $A=(a_{ij})_{n\times m}$ and  $B=(b_{ij})_{n\times m}$ be two matrices. We define $A+B=(a_{ij}+b_{ij})_{n\times m}$ for $1 \leq i\leq n,1\leq j\leq m$.
\end{definition}

\section{Incremental approaches to computing approximations of sets}

In this section, we present incremental
approaches to computing the second and fifth lower and upper
approximations of sets when revising attribute values.

\begin{definition}(Dynamic covering approximation space)
Let $(U,\mathscr{C})$ and $(U,\mathscr{C}^{\ast})$ be covering
approximation spaces, where $U$=$\{x_{1},x_{2},...,x_{n}\}$,
$\mathscr{C}$=$\{C_{1},C_{2},...,C_{m}\}$,
$\mathscr{C}^{\ast}$=$\{C^{\ast}_{1}$,$C^{\ast}_{2},...,C^{\ast}_{m}\}$,
and $C^{\ast}_{i}$=$C_{i}-\{x_{k}\}$ or
$C^{\ast}_{i}$=$C_{i}\cup\{x_{k}\}$ when revising the attribute value
of $x_{k}\in U$. Then $(U,\mathscr{C}^{\ast})$ is
called a dynamic covering approximation space. In addition,  $\mathscr{C}^{\ast}$ is called a dynamic covering.
\end{definition}

In practice, revising attribute values will result in
$|\mathscr{C}^{\ast}|<|\mathscr{C}|$,
$|\mathscr{C}^{\ast}|=|\mathscr{C}|$ and
$|\mathscr{C}^{\ast}|>|\mathscr{C}|$. In this work, we
only discuss the situation that $|\mathscr{C}^{\ast}|=|\mathscr{C}|$
when revising attribute values of an object.

Below, we discuss the relationship between
$\Gamma(\mathscr{C})$ and $\Gamma(\mathscr{C}^{\ast})$. For
convenience, we denote $M_{\mathscr{C}}=(a_{ij})_{n\times m}$,
$M_{\mathscr{C}^{\ast}}=(b_{ij})_{n\times m}$,
$\Gamma(\mathscr{C})=(c_{ij})_{n\times n}$ and
$\Gamma(\mathscr{C}^{\ast})=(d_{ij})_{n\times n}$.

\begin{theorem}
Let $(U,\mathscr{C}^{\ast})$ be a dynamic covering approximation space
of $(U,\mathscr{C})$, $\Gamma(\mathscr{C})$ and
$\Gamma(\mathscr{C}^{\ast})$ the type-1 characteristic matrices of
$\mathscr{C}$ and $\mathscr{C}^{\ast}$, respectively. Then
 \begin{eqnarray*}
 \Gamma(\mathscr{C}^{\ast})=\Gamma(\mathscr{C})+\Delta\Gamma(\mathscr{C})
\end{eqnarray*} where
  \begin{eqnarray*}
  \Delta\Gamma(\mathscr{C})&=&
 \left[\begin{array}{cccccc}
  0&0&\cdots&d^{\ast}_{1k}&...& 0 \\
  0&0&\cdots&d^{\ast}_{2k}&...& 0 \\
  \cdots&\cdots&\cdots&\cdots&\cdots& \cdots \\
  d^{\ast}_{k1}&d^{\ast}_{k2}&\cdots&d^{\ast}_{kk}&\cdots& d^{\ast}_{kn} \\
  \cdots&\cdots&\cdots&\cdots&\cdots& \cdots \\
  0&0&\cdots&d^{\ast}_{nk}&\cdots& 0 \\
  \end{array}
\right];\\
d^{\ast}_{kj}&=&d^{\ast}_{jk}=\left[\begin{array}{cccccc}
  b_{k1}&b_{k2}&\cdots& b_{km} \\
  \end{array}
\right]\cdot \left[\begin{array}{cccccc}
  b_{1j}&b_{2j}&\cdots& b_{mj} \\
  \end{array}
\right]^{T}-c_{kj}.
\end{eqnarray*}
\end{theorem}

\noindent\textbf{Proof.} By Definition 2.4, $\Gamma(\mathscr{C})$
and $\Gamma(\mathscr{C}^{\ast})$ are presented as follows:
\begin{eqnarray*}
\Gamma(\mathscr{C})&=&M_{\mathscr{C}}\cdot
M_{\mathscr{C}}^{T}\\&=&\left[
  \begin{array}{cccc}
    a_{11} & a_{12} & \cdots& a_{1m} \\
    a_{21} & a_{22} & \cdots& a_{2m} \\
    \cdots & \cdots & \cdots & \cdots \\
    a_{n1} & a_{n2} & \cdots & a_{nm} \\
  \end{array}
\right] \cdot \left[
  \begin{array}{cccc}
    a_{11} & a_{12} & \cdots & a_{1m} \\
    a_{21} & a_{22} & \cdots & a_{2m} \\
    \cdots & \cdots & \cdots & \cdots \\
    a_{n1} & a_{n2} & \cdots & a_{nm} \\
  \end{array}
\right]^{T}
\\ &=&\left[
  \begin{array}{cccccc}
    c_{11} & c_{12} & \cdots & c_{1n} \\
    c_{21} & c_{22} & \cdots & c_{2n} \\
    \cdots & \cdots & \cdots & \cdots \\
    c_{n1} & c_{n2} & \cdots & c_{nn} \\
  \end{array}
\right];
\\
\Gamma(\mathscr{C}^{\ast})&=&M_{\mathscr{C}^{\ast}}\cdot
M_{\mathscr{C}^{\ast}}^{T}\\&=&\left[
  \begin{array}{cccccc}
    b_{11} & b_{12} & \cdots & b_{1m} \\
    b_{21} & b_{22} & \cdots & b_{2m} \\
    \cdots & \cdots & \cdots & \cdots \\
    b_{n1} & b_{n2} & \cdots & b_{nm} \\
  \end{array}
\right]\cdot \left[
  \begin{array}{cccccc}
    b_{11} & b_{12} & \cdots & b_{1m} \\
    b_{21} & b_{22} & \cdots & b_{2m} \\
    \cdots & \cdots & \cdots & \cdots \\
    b_{n1} & b_{n2} & \cdots & b_{nm} \\
  \end{array}
\right]^{T}
\\
&=&\left[
  \begin{array}{ccccccc}
    d_{11} & d_{12} & \cdots & d_{1n}\\
    d_{21} & d_{22} & \cdots & d_{2n}\\
    \cdots & \cdots & \cdots & \cdots \\
    d_{n1} & d_{n2} & \cdots & d_{nn}\\
  \end{array}
\right].
\end{eqnarray*}

By Definition 2.4, we have $c_{ij}=d_{ij}$ for $i\neq
k,j\neq k$  since $a_{ij}=b_{ij}$ for $i\neq k$. To compute
$\Gamma(\mathscr{C}^{\ast})$ on the basis of $\Gamma(\mathscr{C})$,
we only need to compute $(d_{ij})_{(i=k, 1\leq j\leq n)}$ and
$(d_{ij})_{(1\leq i\leq n, j=k)}$. Since
$\Gamma(\mathscr{C}^{\ast})$ is symmetric, we only need to compute
$(d_{ij})_{(i=k, 1\leq j\leq n)}$. In other words, we need to
compute $\Delta\Gamma(\mathscr{C})$, where
\begin{eqnarray*}
  \Delta\Gamma(\mathscr{C})&=&
 \left[\begin{array}{cccccc}
  0&0&\cdots&d^{\ast}_{1k}&...& 0 \\
  0&0&\cdots&d^{\ast}_{2k}&...& 0 \\
  \cdots&\cdots&\cdots&\cdots&\cdots& \cdots \\
  d^{\ast}_{k1}&d^{\ast}_{k2}&\cdots&d^{\ast}_{kk}&\cdots& d^{\ast}_{kn} \\
  \cdots&\cdots&\cdots&\cdots&\cdots& \cdots \\
  0&0&\cdots&d^{\ast}_{nk}&\cdots& 0 \\
  \end{array}
\right];
\\d^{\ast}_{kj}&=&d^{\ast}_{jk}=\left[\begin{array}{cccccc}
  b_{k1}&b_{k2}&\cdots& b_{km} \\
  \end{array}
\right]\cdot \left[\begin{array}{cccccc}
  b_{1j}&b_{2j}&\cdots& b_{mj} \\
  \end{array}
\right]^{T}-c_{kj}.
\end{eqnarray*}

Therefore, we have that
$$
\Gamma(\mathscr{C}^{\ast})=\Gamma(\mathscr{C})+\Delta\Gamma(\mathscr{C}). \Box
$$

The following example is employed to show the process of
constructing approximations of sets by Theorem 3.2.

\begin{example}\label{ex:dynamictype1}
Let $U=\{x_{1},x_{2},x_{3},x_{4}\}$,
$\mathscr{C}=\{C_{1},C_{2},C_{3}\}$ and
$\mathscr{C}^{\ast}=\{C^{\ast}_{1},C^{\ast}_{2},C^{\ast}_{3}\}$,
where $C_{1}=\{x_{1},x_{4}\}$, $C_{2}=\{x_{1},x_{2},x_{4}\}$,
$C_{3}=\{x_{3},x_{4}\}$, $C^{\ast}_{1}=\{x_{1},x_{3},x_{4}\}$,
$C^{\ast}_{2}=\{x_{1},x_{2},x_{3},x_{4}\}$,
$C^{\ast}_{3}=\{x_{4}\}$, and $X=\{x_{3},x_{4}\}$. By Definition
2.4, we first have that
\begin{eqnarray*}
\Gamma(\mathscr{C})&=&M_{\mathscr{C}}\cdot
M_{\mathscr{C}}^{T}\\&=&(c_{ij})_{4\times 4}\\&=&\left[
\begin{array}{cccc}
1 & 1 & 0 \\
0 & 1 & 0 \\
0 & 0 & 1 \\
1 & 1 & 1 \\
\end{array}
\right] \cdot \left[
\begin{array}{cccc}
1 & 1 & 0 \\
0 & 1 & 0 \\
0 & 0 & 1 \\
1 & 1 & 1 \\
\end{array}
\right]^{T} \\
&=&\left[
\begin{array}{cccc}
1 & 1 & 0 & 1 \\
1 & 1 & 0 & 1 \\
0 & 0 & 1 & 1 \\
1 & 1 & 1 & 1 \\
\end{array}
\right].
\end{eqnarray*}

Second, we denote  $\Gamma(\mathscr{C}^{\ast})=(d_{ij})_{4\times
4}$. By Theorem 3.2, we get that
\begin{eqnarray*}
\left[\begin{array}{cccccc}
  d^{\ast}_{31}&d^{\ast}_{32}&d^{\ast}_{33}& d^{\ast}_{34} \\
  \end{array}
\right]&=&\left[\begin{array}{ccccc}
 1 & 1 & 0 \\
  \end{array}
\right]\cdot M^{T}_{\mathscr{C}^{\ast}}-\left[\begin{array}{cccccc}
  c_{31}&c_{32}&c_{33}&c_{34} \\
  \end{array}
\right]\\
&=&\left[\begin{array}{ccccc}
 1 & 1 & 0 \\
  \end{array}
\right]\cdot
\left[\begin{array}{cccc}
 1 & 0 & 1 & 1\\
 1 & 1 & 1 & 1 \\
 0 & 0 & 0 & 1\\
  \end{array}
\right]-\left[\begin{array}{ccccc}
 0 & 0 & 1 & 1 \\
  \end{array}
\right]\\&=&
\left[\begin{array}{ccccc}
 1 & 1 & 1 & 1 \\
  \end{array}
\right]-\left[\begin{array}{ccccc}
 0 & 0 & 1 & 1 \\
  \end{array}
\right]\\&=&\left[\begin{array}{ccccc}
 1 & 1 & 0 & 0 \\
  \end{array}
\right];\\
\left[\begin{array}{cccccc}
  d^{\ast}_{13}&d^{\ast}_{23}&d^{\ast}_{33}& d^{\ast}_{43} \\
  \end{array}
\right]&=&\left[\begin{array}{cccccc}
  d^{\ast}_{31}&d^{\ast}_{32}&d^{\ast}_{33}& d^{\ast}_{34} \\
  \end{array}
\right].
\end{eqnarray*}

By Theorem 3.2, we have that
\begin{eqnarray*}
\Delta\Gamma(\mathscr{C})&=&\left[\begin{array}{cccccccccc}
  0&0&d^{\ast}_{13}& 0 \\
  0&0&d^{\ast}_{23}& 0 \\
  d^{\ast}_{31}&d^{\ast}_{32}&d^{\ast}_{33}& d^{\ast}_{34} \\
  0&0&d^{\ast}_{43}& 0 \\
  \end{array}
\right]\\&=&\left[
\begin{array}{cccc}
0 & 0 & 1 & 0 \\
0 & 0 & 1 & 0 \\
1 & 1 & 0 & 0 \\
0 & 0 & 0 & 0 \\
\end{array}
\right].
\end{eqnarray*}

Thus, we obtain that
\begin{eqnarray*}
\Gamma(\mathscr{C}^{\ast})&=&\Gamma(\mathscr{C})+\Delta\Gamma(\mathscr{C})\\
&=&\left[
\begin{array}{cccc}
1 & 1 & 0 & 1 \\
1 & 1 & 0 & 1 \\
0 & 0 & 1 & 1 \\
1 & 1 & 1 & 1 \\
\end{array}
\right]+\left[
\begin{array}{cccc}
0 & 0 & 1 & 0 \\
0 & 0 & 1 & 0 \\
1 & 1 & 0 & 0 \\
0 & 0 & 0 & 0 \\
\end{array}
\right]\\&=&\left[
\begin{array}{cccc}
1 & 1 & 1 & 1 \\
1 & 1 & 1 & 1 \\
1 & 1 & 1 & 1 \\
1 & 1 & 1 & 1 \\
\end{array}
\right].
\end{eqnarray*}

By Definition 2.6, we have that
\begin{eqnarray*}
\mathcal
{X}_{SH(X)}&=&\Gamma(\mathscr{C}^{\ast})\cdot \mathcal {X}_{X};\\
&=&\left[
\begin{array}{cccc}
1 & 1 & 1 & 1 \\
1 & 1 & 1 & 1 \\
1 & 1 & 1 & 1 \\
1 & 1 & 1 & 1 \\
\end{array}
\right] \cdot \left[
\begin{array}{c}
0 \\
0 \\
1 \\
1 \\
\end{array}
\right]\\
&=&\left[
\begin{array}{cccccc}
1 & 1 & 1 & 1\\
\end{array}
\right]^{T};
\\
 \mathcal {X}_{SL(X)}&=&\Gamma(\mathscr{C}^{\ast})\odot
\mathcal {X}_{X}\\&=&\left[
\begin{array}{cccc}
1 & 1 & 1 & 1 \\
1 & 1 & 1 & 1 \\
1 & 1 & 1 & 1 \\
1 & 1 & 1 & 1 \\
\end{array}
\right] \odot \left[
\begin{array}{c}
0 \\
0 \\
1 \\
1 \\
\end{array}
\right]\\&=&\left[
\begin{array}{cccccc}
0 & 0 & 0 & 0 \\
\end{array}
\right]^{T}.
\end{eqnarray*}

Therefore, $SH(X)=\{x_{1},x_{2},x_{3},x_{4}\}$ and
$SL(X)=\emptyset$.
\end{example}

In Example 3.3, we only need to compute $\Delta\Gamma(\mathscr{C})$ by
Theorem 3.2. But there is a need to compute all elements in
$\Gamma(\mathscr{C}^{\ast})$ by Definition 2.4. Therefore, the
computing time of the incremental algorithm is less than the
non-incremental algorithm.

Subsequently, we discuss the construction of
$\prod(\mathscr{C}^{\ast})$ based on $\prod(\mathscr{C})$. For
convenience, we denote $\prod(\mathscr{C})=(e_{ij})_{n\times n}$ and
$\prod(\mathscr{C}^{\ast})=(f_{ij})_{n\times n}$.

\begin{theorem}
Let $(U,\mathscr{C}^{\ast})$ be a dynamic covering approximation space
of $(U,\mathscr{C})$, $\prod(\mathscr{C})$ and
$\prod(\mathscr{C}^{\ast})$ the type-2 characteristic matrice of
$\mathscr{C}$ and $\mathscr{C}^{\ast}$, respectively. Then
 \begin{eqnarray*}
 \prod(\mathscr{C}^{\ast})=\prod(\mathscr{C})+\Delta\prod(\mathscr{C})
\end{eqnarray*} where
  \begin{eqnarray*}
 \Delta\prod(\mathscr{C})&=&
 \left[\begin{array}{cccccc}
  0&0&\cdots&f^{\ast}_{1k}&\cdots& 0 \\
  0&0&\cdots&f^{\ast}_{2k}&\cdots& 0 \\
  \cdots&\cdots&\cdots&\cdots&\cdots&\cdots \\
  f^{\ast}_{k1}&f^{\ast}_{k2}&\cdots&f^{\ast}_{kk}&\cdots& f^{\ast}_{kn} \\
  \cdots&\cdots&\cdots&\cdots&\cdots&\cdots \\
  0&0&\cdots&f^{\ast}_{nk}&\cdots& 0 \\
  \end{array}
\right];
\\\left[\begin{array}{cccccc}
  f^{\ast}_{k1}&f^{\ast}_{k2}&\cdots& f^{\ast}_{kn} \\
  \end{array}
\right]&=&\left[\begin{array}{cccccc}
  b_{k1}&b_{k2}&\cdots& b_{km} \\
  \end{array}
\right]\odot
M^{T}_{\mathscr{C}^{\ast}}-\left[\begin{array}{cccccc}
  e_{k1}&e_{k2}&\cdots& e_{kn} \\
  \end{array}
\right];\\\left[\begin{array}{cccccc}
  f^{\ast}_{1k}&f^{\ast}_{2k}&\cdots& f^{\ast}_{nk} \\
  \end{array}
\right]^{T}&=& M_{\mathscr{C}^{\ast}}\odot\left[\begin{array}{cccccc}
  b_{1k}&b_{2k}&\cdots& b_{mk} \\
  \end{array}
\right]^{T}-\left[\begin{array}{cccccc}
  e_{1k}&e_{2k}&\cdots& e_{nk} \\
  \end{array}
\right].
\end{eqnarray*}
\end{theorem}
\noindent\textbf{Proof.} By Definition 2.5, $\prod(\mathscr{C})$ and
$\prod(\mathscr{C}^{\ast})$ are presented as follows:
\begin{eqnarray*}
\prod(\mathscr{C})&=&M_{\mathscr{C}}\odot
M_{\mathscr{C}}^{T}\\&=&\left[
  \begin{array}{cccc}
    a_{11} & a_{12} & \cdots & a_{1m} \\
    a_{21} & a_{22} & \cdots & a_{2m} \\
    \cdots & \cdots & \cdots & \cdots \\
    a_{n1} & a_{n2} & \cdots & a_{nm} \\
  \end{array}
\right] \odot \left[
  \begin{array}{cccc}
    a_{11} & a_{12} & \cdots & a_{1m} \\
    a_{21} & a_{22} & \cdots & a_{2m} \\
    \cdots & \cdots & \cdots & \cdots \\
    a_{n1} & a_{n2} & \cdots & a_{nm} \\
  \end{array}
\right]^{T}
\\ &=&\left[
  \begin{array}{cccc}
    e_{11} & e_{12} & \cdots & e_{1n} \\
    e_{21} & e_{22} & \cdots & e_{2n} \\
    \cdots & \cdots & \cdots & \cdots \\
    e_{n1} & e_{n2} & \cdots & e_{nn} \\
  \end{array}
\right];
\\
\prod(\mathscr{C}^{\ast})&=&M_{\mathscr{C}^{\ast}}\odot
M_{\mathscr{C}^{\ast}}^{T}\\&=&\left[
  \begin{array}{cccc}
    b_{11} & b_{12} & \cdots & b_{1m} \\
    b_{21} & b_{22} & \cdots & b_{2m} \\
    \cdots & \cdots & \cdots & \cdots \\
    b_{n1} & b_{n2} & \cdots & b_{nm} \\
  \end{array}
\right]\odot \left[
  \begin{array}{cccc}
    b_{11} & b_{12} & \cdots & b_{1m} \\
    b_{21} & b_{22} & \cdots & b_{2m} \\
    \cdots & \cdots & \cdots & \cdots \\
    b_{n1} & b_{n2} & \cdots & b_{nm} \\
  \end{array}
\right]^{T} \\ &=&\left[
  \begin{array}{cccc}
    f_{11} & f_{12} & \cdots & f_{1n}\\
    f_{21} & f_{22} & \cdots & f_{2n}\\
    \cdots & \cdots & \cdots & \cdots \\
    f_{n1} & f_{n2} & \cdots& f_{nn}\\
  \end{array}
\right].
\end{eqnarray*}

By Definition 2.5, we have $e_{ij}=f_{ij}$ for $i\neq
k,j\neq k$ since $a_{ij}=b_{ij}$ for $i\neq k$. To compute
$\prod(\mathscr{C}^{\ast})$ on the basis of $\prod(\mathscr{C})$, we
only need to compute $(f_{ij})_{(i=k, 1\leq j\leq n)}$ and
$(f_{ij})_{(1\leq i\leq n, j=k)}$. In other words, we need to
compute $\Delta\prod(\mathscr{C})$, where
\begin{eqnarray*}
\Delta\prod(\mathscr{C})&=&
 \left[\begin{array}{cccccc}
  0&0&\cdots&f^{\ast}_{1k}&\cdots& 0 \\
  0&0&\cdots&f^{\ast}_{2k}&\cdots& 0 \\
  \cdots&\cdots&\cdots&\cdots&\cdots&\cdots\\
  f^{\ast}_{k1}&f_{k2}&\cdots&f^{\ast}_{kk}&\cdots& f^{\ast}_{kn} \\
  \cdots&\cdots&\cdots&\cdots&\cdots&\cdots\\
  0&0&\cdots&f^{\ast}_{nk}&\cdots& 0 \\
  \end{array}
\right];
\\\left[\begin{array}{cccccc}
  f^{\ast}_{k1}&f^{\ast}_{k2}&\cdots& f^{\ast}_{kn} \\
  \end{array}
\right]&=&\left[\begin{array}{cccccc}
  b_{k1}&b_{k2}&\cdots& b_{km} \\
  \end{array}
\right]\odot
M^{T}_{\mathscr{C}^{\ast}}-\left[\begin{array}{cccccc}
  e_{k1}&e_{k2}&\cdots& e_{kn} \\
  \end{array}
\right];\\\left[\begin{array}{cccccc}
  f^{\ast}_{1k}&f^{\ast}_{2k}&\cdots& f^{\ast}_{nk} \\
  \end{array}
\right]^{T}&=& M_{\mathscr{C}^{\ast}}\odot\left[\begin{array}{cccccc}
  b_{1k}&b_{2k}&\cdots& b_{mk} \\
  \end{array}
\right]^{T}-\left[\begin{array}{cccccc}
  e_{1k}&e_{2k}&\cdots& e_{nk} \\
  \end{array}
\right].
\end{eqnarray*}

Therefore, we have that
\begin{eqnarray*}
\prod(\mathscr{C}^{\ast})=\prod(\mathscr{C})+\Delta\prod(\mathscr{C}).\Box
\end{eqnarray*}

The following example is employed to show the process of
constructing approximations of sets by Theorem 3.4.

\begin{example}
Let $U=\{x_{1},x_{2},x_{3},x_{4}\}$,
$\mathscr{C}=\{C_{1},C_{2},C_{3}\}$ and
$\mathscr{C}^{\ast}=\{C^{\ast}_{1},C^{\ast}_{2},C^{\ast}_{3}\}$,
where $C_{1}=\{x_{1},x_{4}\}$, $C_{2}=\{x_{1},x_{2},x_{4}\}$,
$C_{3}=\{x_{3},x_{4}\}$, $C^{\ast}_{1}=\{x_{1},x_{3},x_{4}\}$,
$C^{\ast}_{2}=\{x_{1},x_{2},x_{3},x_{4}\}$,
$C^{\ast}_{3}=\{x_{4}\}$, and $X=\{x_{3},x_{4}\}$. By
Definition 2.5, we first have that
\begin{eqnarray*}
\prod(\mathscr{C})&=&M_{\mathscr{C}}\odot
M_{\mathscr{C}}^{T}\\&=&(e_{ij})_{4\times 4}\\&=&\left[
\begin{array}{cccc}
1 & 1 & 0 \\
0 & 1 & 0 \\
0 & 0 & 1 \\
1 & 1 & 1 \\
\end{array}
\right] \odot \left[
\begin{array}{cccc}
1 & 1 & 0 \\
0 & 1 & 0 \\
0 & 0 & 1 \\
1 & 1 & 1 \\
\end{array}
\right]^{T} \\
&=&\left[
\begin{array}{cccc}
1 & 0 & 0 & 1 \\
1 & 1 & 0 & 1 \\
0 & 0 & 1 & 1 \\
0 & 0 & 0 & 1 \\
\end{array}
\right].
\end{eqnarray*}

Second, we denote  $\prod(\mathscr{C}^{\ast})=(f_{ij})_{4\times 4}$.
By Theorem 3.4, we get that
\begin{eqnarray*}
\left[\begin{array}{cccccc}
  f^{\ast}_{31}&f^{\ast}_{32}&f^{\ast}_{33}& f^{\ast}_{34} \\
  \end{array}
\right]&=&\left[\begin{array}{ccccc}
 1 & 1 & 0 \\
  \end{array}
\right]\odot M^{T}_{\mathscr{C}^{\ast}}-\left[\begin{array}{cccccc}
  e_{31}&e_{32}&e_{33}&e_{34} \\
  \end{array}
\right]\\
&=&\left[\begin{array}{ccccc}
1 & 1 & 0 \\
  \end{array}
\right]\odot \left[\begin{array}{ccccc}
 1 & 1 & 0 \\
 0 & 1 & 0 \\
 1 & 1 & 0 \\
 1 & 1 & 1 \\
  \end{array}
\right]^{T}-\left[\begin{array}{ccccc}
 0 & 0 & 1 & 1 \\
  \end{array}
\right]\\&=& \left[\begin{array}{ccccc}
 1 & 0 & 1 & 1 \\
  \end{array}
\right]-\left[\begin{array}{ccccc}
 0 & 0 & 1 & 1 \\
  \end{array}
\right]\\&=&\left[\begin{array}{ccccc}
 1 & 0 & 0 & 0 \\
  \end{array}
\right];\\
\left[\begin{array}{cccccc}
  f^{\ast}_{13}&f^{\ast}_{23}&f^{\ast}_{33}& f^{\ast}_{43} \\
  \end{array}
\right]&=& \left[\begin{array}{ccccc}
 1 & 1 & 0 \\
 0 & 1 & 0 \\
 1 & 1 & 0 \\
 1 & 1 & 1 \\
  \end{array}
\right]\odot \left[\begin{array}{cccc}
 1 & 1 & 0 \\
  \end{array}
\right]^{T} -\left[\begin{array}{cccccc}
  e_{13}&e_{23}&e_{33}& e_{43} \\
  \end{array}
\right]\\&=&\left[\begin{array}{ccccc}
 1 & 1 & 1 & 0 \\
  \end{array}
\right]^{T}-\left[\begin{array}{cccccc}
 0 & 0 & 1 & 0 \\
  \end{array}
\right]\\&=& \left[\begin{array}{ccccc}
 1 & 1 & 0 & 0 \\
  \end{array}
\right]^{T}.
\end{eqnarray*}

By Theorem 3.4, we have that
\begin{eqnarray*}
\Delta\prod(\mathscr{C})&=&\left[\begin{array}{cccccccccc}
  0&0&f^{\ast}_{13}& 0 \\
  0&0&f^{\ast}_{23}& 0 \\
  f^{\ast}_{31}&f^{\ast}_{32}&f^{\ast}_{33}& f^{\ast}_{34} \\
  0&0&f^{\ast}_{43}&. 0 \\
  \end{array}
\right]\\&=&\left[
\begin{array}{cccc}
0 & 0 & 1 & 0 \\
0 & 0 & 1 & 0 \\
1 & 0 & 0 & 0 \\
0 & 0 & 0 & 0 \\
\end{array}
\right].
\end{eqnarray*}

Therefore, we obtain that
\begin{eqnarray*}
\prod(\mathscr{C}^{\ast})&=&\prod(\mathscr{C})+\Delta\prod(\mathscr{C})\\
&=&\left[
\begin{array}{cccc}
1 & 0 & 0 & 1 \\
1 & 1 & 0 & 1 \\
0 & 0 & 1 & 1 \\
0 & 0 & 0 & 1 \\
\end{array}
\right]+\left[
\begin{array}{cccc}
0 & 0 & 1 & 0 \\
0 & 0 & 1 & 0 \\
1 & 0 & 0 & 0 \\
0 & 0 & 0 & 0 \\
\end{array}
\right]\\&=&\left[
\begin{array}{cccc}
1 & 0 & 1 & 1 \\
1 & 1 & 1 & 1 \\
1 & 0 & 1 & 1 \\
0 & 0 & 0 & 1 \\
\end{array}
\right].
\end{eqnarray*}

By Definition 2.6, we have that
\begin{eqnarray*}
\mathcal
{X}_{SH(X)}&=&\prod(\mathscr{C}^{\ast})\cdot \mathcal {X}_{X};\\
&=&\left[
\begin{array}{cccc}
1 & 0 & 1 & 1 \\
1 & 1 & 1 & 1 \\
1 & 0 & 1 & 1 \\
0 & 0 & 0 & 1 \\
\end{array}
\right] \cdot \left[
\begin{array}{c}
0 \\
0 \\
1 \\
1 \\
\end{array}
\right]\\
&=&\left[
\begin{array}{cccccc}
1 & 1 & 1 & 1\\
\end{array}
\right]^{T};
\\
 \mathcal {X}_{SL(X)}&=&\prod(\mathscr{C}^{\ast})\odot
\mathcal {X}_{X}\\&=&\left[
\begin{array}{cccc}
1 & 0 & 1 & 1 \\
1 & 1 & 1 & 1 \\
1 & 0 & 1 & 1 \\
0 & 0 & 0 & 1 \\
\end{array}
\right] \odot \left[
\begin{array}{c}
0 \\
0 \\
1 \\
1 \\
\end{array}
\right]\\&=&\left[
\begin{array}{cccccc}
0 & 0 & 0 & 1 \\
\end{array}
\right]^{T}.
\end{eqnarray*}

Therefore, $SH(X)=\{x_{1},x_{2},x_{3},x_{4}\}$ and
$SL(X)=\{x_{4}\}$.
\end{example}

In Example 3.5, we only need to $\Delta\prod(\mathscr{C})$ by
Theorem 3.4. But there is a need to compute all elements in
$\prod(\mathscr{C}^{\ast})$ by Definition 2.5. Therefore, the
computing time of the incremental algorithm is less than the
non-incremental algorithm.

\section{Non-incremental and incremental algorithms of computing approximations of sets with varying attribute values}

In this section, we present non-incremental and incremental algorithms of computing the second and sixth lower and upper approximations of sets with varying attribute values.

\begin{algorithm}[!h]
\caption{Non-incremental algorithm of computing the second lower and upper approximations of sets(NIS)}
\label{algo:static1}
\KwIn{$(U,\mathscr{C}^{\ast})$
and $X\subseteq U$.	}
\KwOut{$\mathcal {X}_{SH(X)}$ and $\mathcal {X}_{SL(X)}$.}
\Begin
{
	Construct $M_{\mathscr{C}^{\ast}}$ based on $\mathscr{C}^{\ast}$;
	
	Compute
	$\Gamma(\mathscr{C}^{\ast})=M_{\mathscr{C}}\cdot M^{T}_{\mathscr{C}^{\ast}};$
	
	Obtain $\mathcal
		{X}_{SH(X)}=\Gamma(\mathscr{C}^{\ast})\cdot \mathcal {X}_{X}$ and
$\mathcal {X}_{SL(X)}=\Gamma(\mathscr{C}^{\ast})\odot \mathcal {X}_{X}.$	
}
\end{algorithm}

\begin{algorithm}[!h]
\caption{Incremental algorithm of computing the second lower and upper approximations of sets(IS)}
\label{algo:dynamic1}
\KwIn{1. $(U,\mathscr{C})$,
$\Gamma(\mathscr{C})$,
$(U,\mathscr{C}^{\ast})$,
$X\subseteq U$.	}
\KwOut{$\mathcal {X}_{SH(X)}$ and $\mathcal {X}_{SL(X)}$.}
\Begin
{
	Construct $M_{\mathscr{C}}^{\ast}$=$(b_{ij})_{n\times m}$ based on $\mathscr{C}^{\ast}$;
	
	Denote $row_k$=$[b_{k1}, b_{k2}, ...,b_{km}]$;
	
	Compute $\Delta row_k$=$row_k \cdot M^{T}_{\mathscr{C}^{\ast}}$;
	
	Let $\Gamma(\mathscr{C}^{\ast})$=$\Gamma(\mathscr{C})$;
	
	 Set  $k$th row of $\Gamma(\mathscr{C}^{\ast})$ as $\Delta row_k$;
	
	 Set  $k$th col of $\Gamma(\mathscr{C}^{\ast})$ as $(\Delta row_k)^T$;
		
	Obtain
		$\mathcal{X}_{SH(X)}$=$\Gamma(\mathscr{C}^{\ast})\cdot \mathcal {X}_{X}$ and
	$\mathcal{X}_{SL(X)}$=$\Gamma(\mathscr{C}^{\ast})\odot \mathcal
		{X}_{X}$.
	
}
\end{algorithm}

In  Algorithm 4.1, the time complexity of Step 3 is $O(mn^{2})$; the time complexity of step 4 is $O(2n^{2})$. The total time complexity is $O((m+2)n^{2})$.
 In Algorithm 4.2, the time complexity of  Step 4 is $O(nm)$; the time complexity of Step 6 is $O(n)$; the time complexity of Step 7 is $O(n)$; the time complexity of Step 8 is $O(2n^{2})$. The total time complexity is $O(2n^{2}+nm+2n)$. Furthermore, $O((m+2)n^{2})$ is the time
complexity of the non-incremental algorithm. Thus the incremental
algorithm is more effective than the non-incremental algorithm.

\begin{algorithm}[!h]
\caption{Non-incremental algorithm of computing the sixth lower and upper approximations of sets(NIX)}
\label{algo:static2}
\KwIn{$(U,\mathscr{C}^{\ast})$ and
	$X\subseteq U$.	}
\KwOut{$\mathcal {X}_{XH(X)}$ and $\mathcal {X}_{XL(X)}$.}
\Begin
{
	Construct $M_{\mathscr{C}^{\ast}}$ based on $\mathscr{C}^{\ast}$;
	
	Compute
	$\prod(\mathscr{C}^{\ast})=M_{\mathscr{C}^{\ast}}\odot M^{T}_{\mathscr{C}^{\ast}};$
	
	Obtain $\mathcal
	{X}_{XH(X)}=\prod(\mathscr{C}^{\ast})\cdot \mathcal {X}_{X}$ and $\mathcal {X}_{XL(X)}=\prod(\mathscr{C}^{\ast})\odot \mathcal {X}_{X}.$	
}
\end{algorithm}

\begin{algorithm}[!h]
	\caption{Incremental algorithm of computing the sixth lower and upper approximations of sets(IX)}
	\label{algo:dynamic2}
	\KwIn{$(U,\mathscr{C})$,
		$\prod(\mathscr{C})$,
		$(U,\mathscr{C}^{\ast})$ and
		$X\subseteq U$.	}
	\KwOut{$\mathcal {X}_{XH(X)}$ and $\mathcal {X}_{XL(X)}$.}
	\Begin
	{
		Construct $M_{\mathscr{C}}^{\ast}$=$(b_{ij})_{n\times m}$ based on $\mathscr{C}^{\ast}$;
		
		Denote $row_k$=$[b_{k1}, b_{k2}, ...,b_{km}]$;
		
		Compute $\Delta row_k$=$row_k \odot M^{T}_{\mathscr{C}^{\ast}}$;

		Denote $col_k$=$[b_{1k}, b_{2k}, ...,b_{mk}]^T$;
		
		Compute $\Delta col_k$=$M_{\mathscr{C}^{\ast}} \odot col_k$;
				
		Let $\prod(\mathscr{C}^{\ast})$=$\prod(\mathscr{C})$;
		
		Set  $k$th row of $\prod(\mathscr{C}^{\ast})$ as $\Delta row_k$;
		
		Set  $k$th col of $\prod(\mathscr{C}^{\ast})$ as $\Delta col_k$;
		
		Obtain
		$\mathcal{X}_{XH(X)}$=$\prod(\mathscr{C}^{\ast})\cdot \mathcal {X}_{X}$ and $\mathcal{X}_{XL(X)}$=$\prod(\mathscr{C}^{\ast})\odot \mathcal
		{X}_{X}$.
		
	}
\end{algorithm}

In  Algorithm 4.3, the time complexity of Step 3 is $O(mn^{2})$, the time complexity of Step 4 is $O(n^{2})$. The total time complexity is $O((m+2)n^{2})$. In Algorithm 4.4, the time complexity of  Step 4 is $O(nm)$; the time complexity of Step 6 is $O(nm)$; the time complexity of Step 8 is $O(n)$; the time complexity of Step 9 is $O(n)$; the time complexity of Step 10 is $O(2n^{2})$. The total time complexity is $O(2n^{2}+2nm+2n)$. Furthermore, $O((m+2)n^{2})$ is the time
complexity of the non-incremental algorithm. Thus the incremental
algorithm is more effective than the non-incremental algorithm.

\section{Experimental analysis}

In this section, we perform the series of experiments to validate the effectiveness of Algorithms 4.2 and 4.4 for computing approximations in dynamic covering approximation spaces when varying attribute values.

\subsection{Experimental environment}

 Since transforming information systems into covering approximation spaces takes a great deal of time, and the main objective of this work is to illustrate the efficiency of the Algorithms 4.2 and 4.4 in computing approximations of sets. To evaluate the performance of Algorithms 4.2 and 4.4,  we generated ten covering approximation spaces $(U_{i},\mathscr{C}_{i})$ for the experiment, where $i,j=1,2,3,...,10.$
We outline all these covering approximation spaces in Table 1, where $|U_{i}|$ denotes the number of objects in $U_{i}$ and $|\mathscr{C}_{i}|$ is the cardinality of $\mathscr{C}_{i}$.

All computations were conducted on a PC with a Inter(R) Core(TM) i5-4200M CPU $@$ 2.50 GHZ and 4 GB memory, running 64-bit Windows 7 Service Pack 1. The software used was 64-bit Matlab R2013b. Details of the hardware and software are given in Table 2.

\begin{table}\renewcommand{\arraystretch}{1.5}
\caption{Covering approximation spaces.
} \tabcolsep0.57in
\begin{tabular}{cccc}
\hline
No.&Name&$|U_{i}|$ &$|\mathscr{C}_{i}|$\\\hline
1  & $(U_{1},\mathscr{C}_{1})$ &2000& 100\\
2  & $(U_{2},\mathscr{C}_{2})$ &4000& 200\\
3  & $(U_{3},\mathscr{C}_{3})$ &6000& 300\\
4  & $(U_{4},\mathscr{C}_{4})$ &8000& 400\\
5  & $(U_{5},\mathscr{C}_{5})$ &10000& 500\\
6  & $(U_{6},\mathscr{C}_{6})$ &12000& 600\\
7  & $(U_{7},\mathscr{C}_{7})$ &14000& 700\\
8  & $(U_{8},\mathscr{C}_{8})$ &16000& 800\\
9  & $(U_{9},\mathscr{C}_{9})$ &18000& 900\\
10  & $(U_{10},\mathscr{C}_{10})$ &20000& 1000\\
\hline
\end{tabular}
\end{table}

\begin{table}\renewcommand{\arraystretch}{1.5}
\caption{The experimental environment.
} \tabcolsep0.35in
\begin{tabular}{cccc}
\hline
No.&Name&Model &Parameters\\\hline
1  & CPU & Inter(R) Core(TM) i5-4200M  & 2.50 GHZ \\
2 & Memory  & ADAT DDR3 &4G \\
3  & Hard disk &SATA& 1T\\
4 & System  &Windows 7 & 64bit \\
5  & Platform &Matlab R2013b& 64bit\\
\hline
\end{tabular}
\end{table}

\subsection{Experimental results}

\subsubsection{Computational times in dynamic covering approximation spaces}

In this subsection, we apply Algorithms 4.1-4.4 to the covering
approximation space $(U_{i},\mathscr{C}_{i})$, where $i=1,2,3,...,10$, and compare the computing times by using Algorithms 4.1 and 4.3 with those of Algorithms 4.2 and 4.4, respectively.

First, we calculate $\Gamma(\mathscr{C}_{i})$ and
$\prod(\mathscr{C}_{i})$ by Definitions 2.4 and 2.5, respectively.
We also obtain the dynamic covering approximation space
$(U_{i},\mathscr{C}^{\ast}_{i})$ when revising attribute values of $x_{k}$, where
and $C^{\ast}_{j}=C_{j}\cup \{x_{k}\}$ or $C_{j}$, where $C^{\ast}_{j}\in\mathscr{C}^{\ast}_{i}$ and $C_{j}\in\mathscr{C}_{i}$.
Subsequently, we get $\Gamma(\mathscr{C}^{\ast}_{i})$ and
$\prod(\mathscr{C}^{\ast}_{i})$ by Algorithms 4.1 and 4.3,
respectively.

Second, we calculate $SH(X)$, $SL(X)$, $ XH(X)$ and $XL(X)$ based on $\Gamma(\mathscr{C}^{\ast}_{i})$ and $
\prod(\mathscr{C}^{\ast}_{i})$ for $X\subseteq U_{i}$,
respectively. The time of computing $SH(X)$, $SL(X)$,
$XH(X)$ and $XL(X)$ is shown in Tables 3-12. Concretely, $NIS$
and  $NIX$ stands for the time of constructing
the second and sixth lower and upper approximations of sets by Algorithms 4.1 and 4.3 in Tables 3-12.
Additionally, we obtain $\Gamma(\mathscr{C}^{\ast}_{i})$ and
$\prod(\mathscr{C}^{\ast}_{i})$ by Algorithms 4.2 and 4.4,
respectively. Then the time of computing $SH(X)$, $SL(X)$,
$XH(X)$ and $XL(X)$ for $X\subseteq U_{i}$ is shown in Tables 3-12.
Concretely, $IS$
and  $IX$ stands for the time of computing
the second and sixth lower and upper approximations of sets by Algorithms 4.2 and 4.4 in Tables 3-12.

Third, we conduct all experiments ten times and show the results in
Tables 3-12 and Figures 1-10. We see all
algorithms are stable to compute approximations of sets in all experiments. Concretely, we observe that the computing times by using the same
algorithm are almost the same in Tables 3-12. Consequently, we see that
the times of computing approximations of sets by using incremental
algorithms are much smaller than those of the non-incremental
algorithms.  In Figures 1-10, we also observe that the computing times
of Algorithms 4.2 and 4.4 are far less than those of Algorithms 4.1 and 4.3, respectively. Therefore, the incremental algorithms
are more effective to construct approximations of sets in the
dynamic covering approximation space
$(U_{i},\mathscr{C}^{\ast}_{i})$, where $i=1,2, ..., 10.$

\noindent\textbf{Remark:}  In Tables 3-12, the
measure of time is in seconds; $\overline{t}$ indicates the average time of
ten experiments; In Figures
1-10, $i$ stands for the experimental number in $X$
Axis; In Figure 11, $i$ refers to as the covering approximation space $(U_{i},\mathscr{C}_{i})$ in $X$
Axis; In Figures 1-11, $i$ is the computing time
in $Y$ Axis.

\begin{table}\renewcommand{\arraystretch}{1.5}
\caption{Computational times using Algorithms 4.1-4.4 in
$(U_{1},\mathscr{C}_{1})$. } \tabcolsep0.04in
\begin{tabular}{ccccccccccccc}
\hline Algorithm&1 & 2& 3 & 4 & 5& 6& 7& 8 & 9&
10&$\overline{t}$\\\hline
NIS& 0.4578 &    0.4213  &    0.4279   &   0.4223  &    0.4271  &    0.4236 &     0.4235 &     0.4263  &    0.4236 &     0.4273  &    0.4281\\
NIX&     0.4681   &   0.4671  &    0.4636  &    0.4646  &    0.4668  &    0.4651    &  0.4651   &   0.4681   &   0.4668   &   0.4720  &    0.4667\\
IS&  0.0044   &   0.0026  &    0.0033   &   0.0040  &    0.0029  &    0.0028 &     0.0031  &    0.0030   &   0.0028  &    0.0028   &   0.0032\\
IX&  0.0351   &   0.0339   &   0.0333  &    0.0339  &    0.0340  &    0.0334  &    0.0340     & 0.0335  &    0.0338  &    0.0333   &   0.0338\\
\hline
\end{tabular}
\end{table}

\begin{figure}
\begin{center}
\includegraphics[width=8cm]{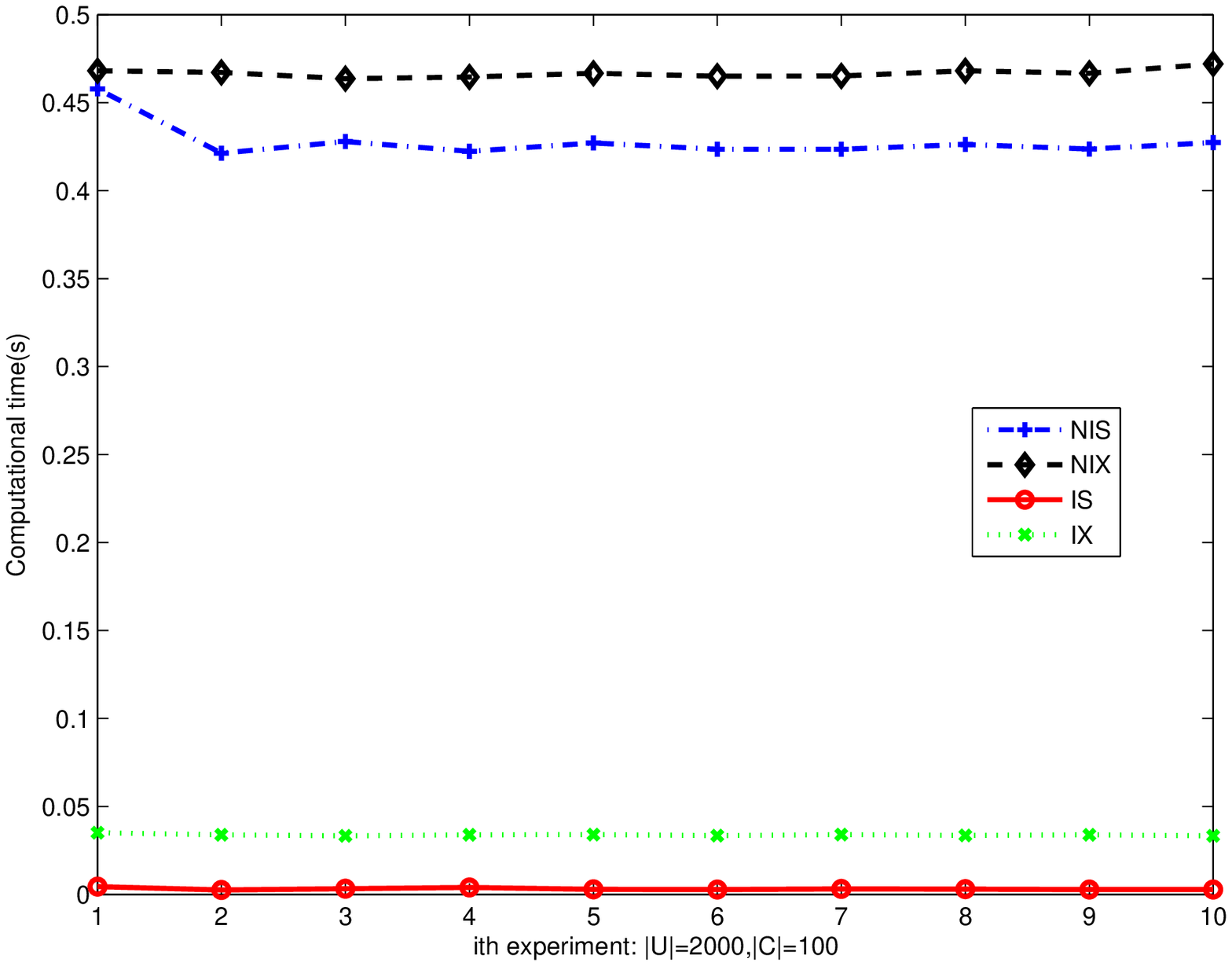}\\
\caption{Computational times using Algorithms 4.1-4.4 in
$(U_{1},\mathscr{C}_{1})$.}
\end{center}
\end{figure}

\begin{table}\renewcommand{\arraystretch}{1.5}
\caption{Computational times using Algorithms 4.1-4.4 in
$(U_{2},\mathscr{C}_{2})$. } \tabcolsep0.04in
\begin{tabular}{ccccccccccccc}
\hline Algorithm&1 & 2& 3 & 4 & 5& 6& 7& 8 & 9&
10&$\overline{t}$\\\hline
 NIS&1.8902  &   1.8452   &  1.8610   &  1.8203  &   1.8179  &   1.8257  &   1.8223  &   1.8224   &  1.8294   &  1.8189    & 1.8353\\
NIX&2.0389   &  2.0437    & 2.0314   &  2.0237  &   2.0378   &  2.0331    & 2.0531&     2.0565   &  2.0583  &   2.0641   &  2.0440\\
IS&0.0091  &   0.0118 &    0.0102  &   0.0100 &    0.0098  &   0.0127 &    0.0110   &  0.0099  &   0.0099 &    0.0096 &    0.0104\\
IX&0.2035  &   0.2018  &   0.2013  &   0.2018  &   0.2034  &   0.1992 &    0.2018  &   0.2006   &  0.1987   &  0.2035   &  0.2016\\
\hline
\end{tabular}
\end{table}

\begin{figure}
\begin{center}
\includegraphics[width=8cm]{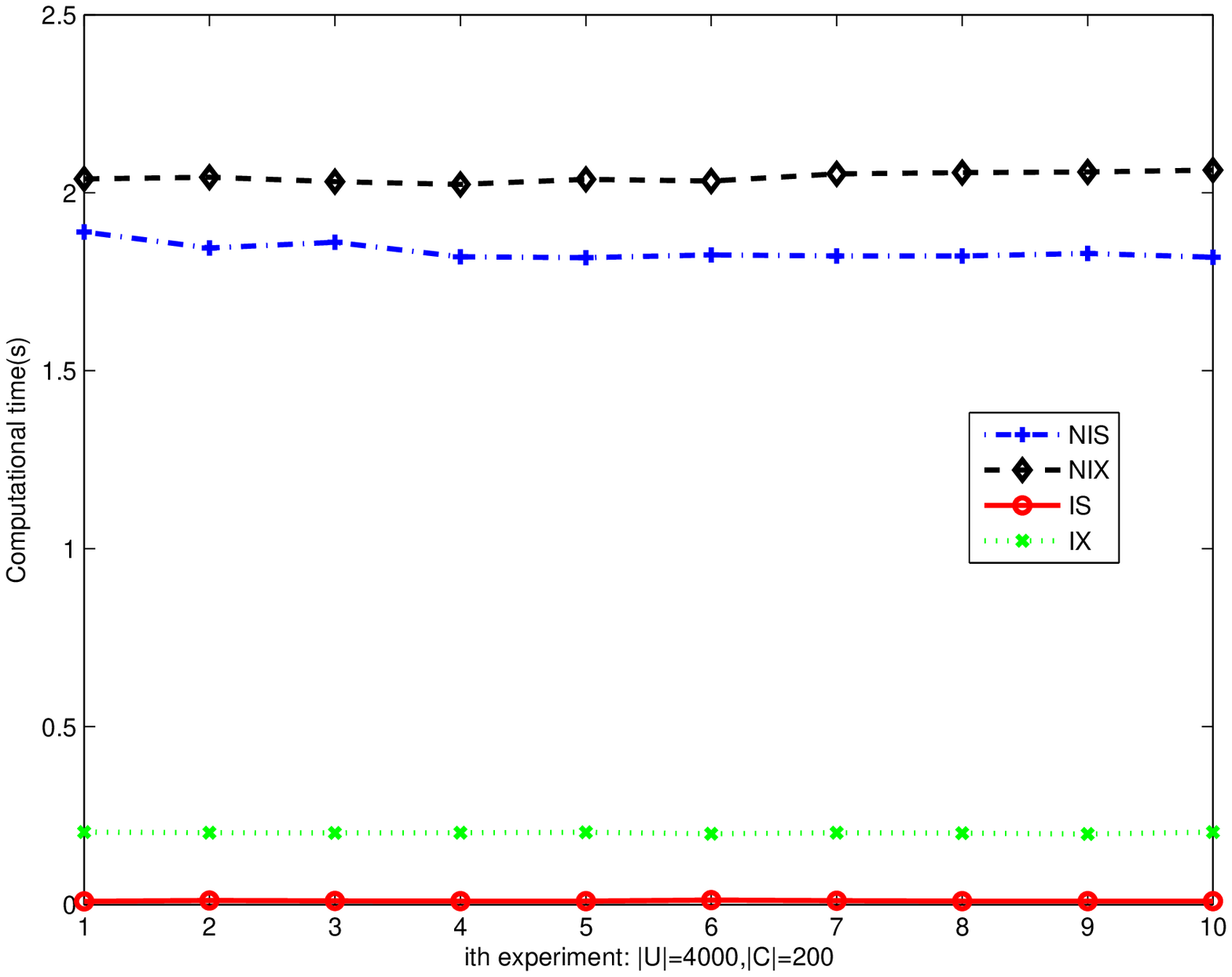}\\
\caption{Computational times using Algorithms 4.1-4.4 in
$(U_{2},\mathscr{C}_{2})$.}
\end{center}
\end{figure}

\begin{table}\renewcommand{\arraystretch}{1.5}
\caption{Computational times using Algorithms 4.1-4.4 in
$(U_{3},\mathscr{C}_{3})$. } \tabcolsep0.04in
\begin{tabular}{ccccccccccccc}
\hline Algorithm&1 & 2& 3 & 4 & 5& 6& 7& 8 & 9&
10&$\overline{t}$\\\hline
NIS&4.2030  &   4.1889  &   4.1905  &   4.1457  &   4.1446  &   4.1681 &    4.1518 &    4.1765   &  4.2310  &   4.1604  &   4.1760\\
NIX&4.6993  &   4.7126  &   4.6838  &   4.6895  &   4.6941  &   4.7000  &   4.7025  &   4.6711  &   4.7039 &    4.6939   &  4.6951\\
IS&0.0177 &    0.0210   &  0.0211 &    0.0199 &    0.0199 &    0.0199  &   0.0199  &   0.0205  &   0.0200 &    0.0197   &  0.0200\\
IX&0.5259  &   0.5059 &    0.5076  &   0.5056 &    0.5089 &    0.5055   &  0.5106 &    0.5080  &   0.5059   &  0.5078 &    0.5092\\
\hline
\end{tabular}
\end{table}

\begin{figure}
\begin{center}
\includegraphics[width=8cm]{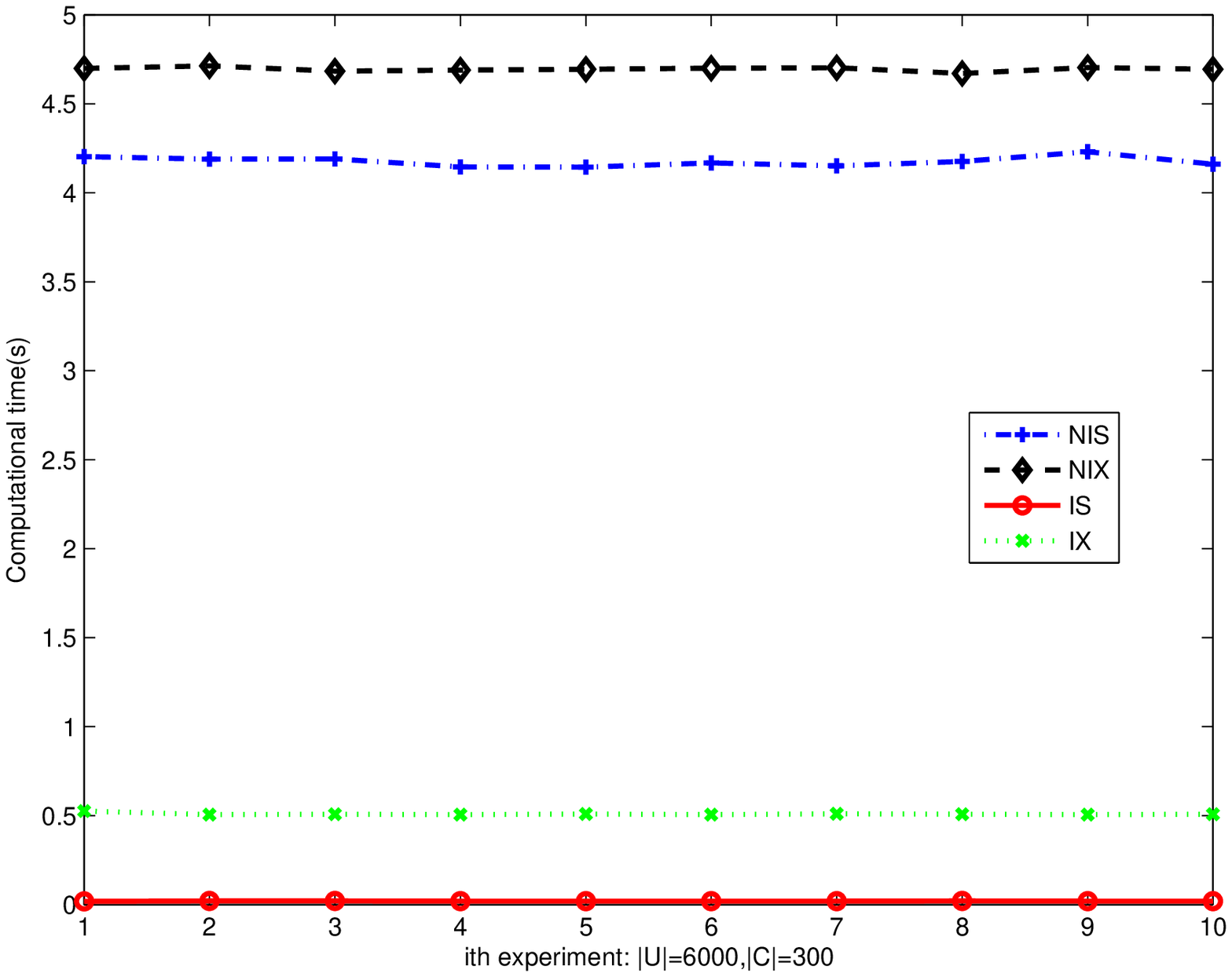}\\
\caption{Computational times using Algorithms 4.1-4.4 in
$(U_{3},\mathscr{C}_{3})$.}
\end{center}
\end{figure}

\begin{table}\renewcommand{\arraystretch}{1.5}
\caption{Computational times using Algorithms 4.1-4.4 in
$(U_{4},\mathscr{C}_{4})$. } \tabcolsep0.04in
\begin{tabular}{cccccccccccc}
\hline Algorithm&1 & 2& 3 & 4 & 5& 6& 7& 8 & 9&
10&$\overline{t}$\\\hline
 NIS&7.5968  &   7.5550  &   7.7428  &   7.6536  &   7.6756  &   7.7031 &    7.6304     &7.6051   &  7.6118  &   7.7013  &   7.6475\\
NIX&8.6892  &   8.7967   &  8.8918  &   9.0384  &   8.7810  &   8.7764  &   8.6300  &   9.2821  &   8.6324  &   8.6121 &    8.8130\\
IS&0.0428  &   0.0338 &    0.0350  &   0.0394  &   0.0378 &    0.0386   &  0.0345&     0.0345  &   0.0346   &  0.0348  &   0.0366\\
IX&0.9813  &   0.9681  &   0.9694  &   0.9677  &   0.9669 &    0.9731  &   0.9654   &  0.9683   &  0.9648  &   0.9685 &    0.9694\\
\hline
\end{tabular}
\end{table}

\begin{figure}
\begin{center}
\includegraphics[width=8cm]{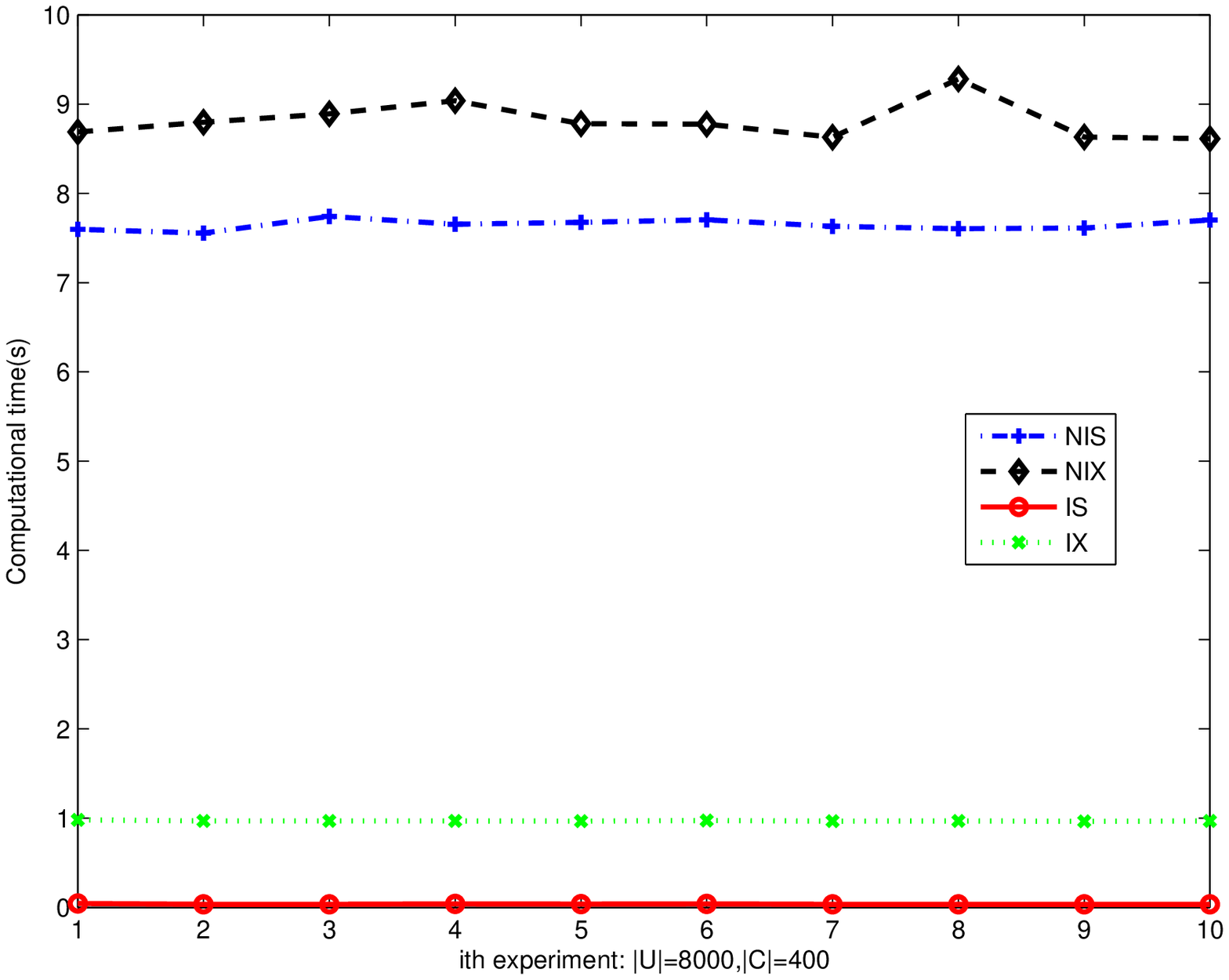}\\
\caption{Computational times using Algorithms 4.1-4.4 in
$(U_{4},\mathscr{C}_{4})$.}
\end{center}
\end{figure}

\begin{table}\renewcommand{\arraystretch}{1.5}
\caption{Computational times using Algorithms 4.1-4.4 in
$(U_{5},\mathscr{C}_{5})$. } \tabcolsep0.01in
\begin{tabular}{cccccccccccc}
\hline Algorithm&1 & 2& 3 & 4 & 5& 6& 7& 8 & 9&
10&$\overline{t}$\\\hline
 NIS&12.0856 &   11.9662&    11.9944  &  11.9200  &  11.9992 &   11.9683  &  11.9321&    11.9008  &  11.8811 &   11.8839  &  11.9532\\
NIX&13.8290  &  13.6560  &  13.7430  &  13.7308  &  13.6831 &   13.6816 &   13.7970  &  13.6794  &  13.8141  &  13.7338  &  13.7348\\
IS&0.0675   &  0.0530  &   0.0549  &   0.0537  &   0.0551  &   0.0537  &   0.0536  &   0.0523   &  0.0535  &   0.0540 &    0.0551\\
IX&1.6266  &   1.6193  &   1.6163  &   1.6138  &   1.6189 &    1.6057  &   1.6230 &    1.6213   &  1.6172  &   1.6211  &   1.6183\\
\hline
\end{tabular}
\end{table}

\begin{figure}
\begin{center}
\includegraphics[width=8cm]{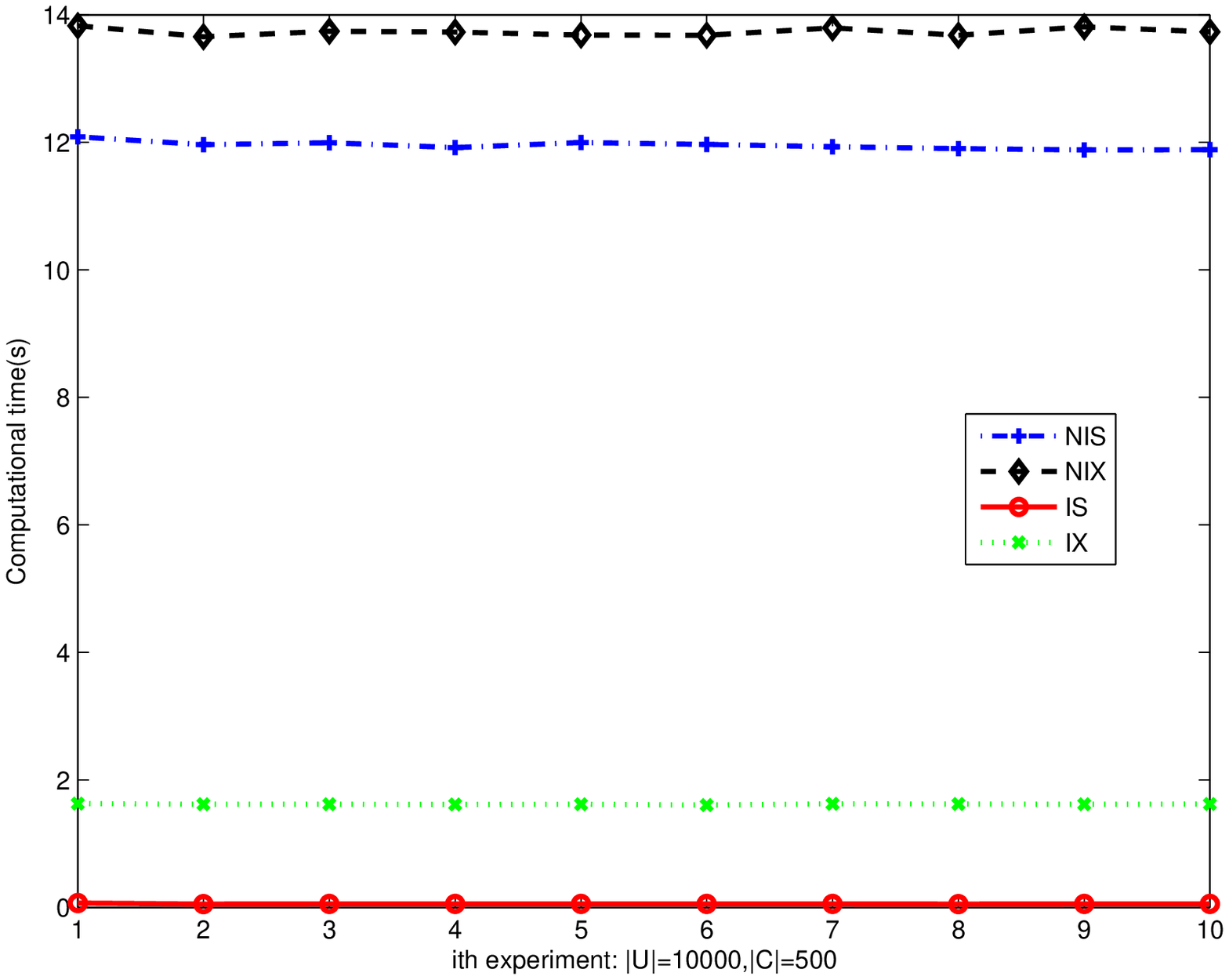}\\
\caption{Computational times using Algorithms 4.1-4.4 in
$(U_{5},\mathscr{C}_{5})$.}
\end{center}
\end{figure}

\begin{table}\renewcommand{\arraystretch}{1.5}
\caption{Computational times using Algorithms 4.1-4.4 in
$(U_{6},\mathscr{C}_{6})$. } \tabcolsep0.01in
\begin{tabular}{cccccccccccc}
\hline Algorithm&1 & 2& 3 & 4 & 5& 6& 7& 8 & 9&
10&$\overline{t}$\\\hline
 NIS&17.8842  &  17.8858 &   18.0800 &   17.6753 &   17.5945&    17.5710 &   17.7019 &   18.2036  &  17.5415  &  17.9582  &  17.8096\\
NIX&20.1684  &  20.1404  &  20.0242  &  20.0022  &  20.0277 &   20.0598  &  20.0897 &   20.2560  &  21.6223  &  22.1194  &  20.4510\\
IS&0.0977  &   0.0748  &   0.0746  &   0.0744  &   0.0803  &   0.0727  &   0.0753  &   0.0735   &  0.0738  &   0.0723  &   0.0770\\
IX&2.4011  &   2.3671   &  2.4204  &   2.3771 &    2.3679  &   2.3662  &   2.3737  &   2.3644  &   2.3614   &  2.3692  &   2.3769
\\
\hline
\end{tabular}
\end{table}

\begin{figure}
\begin{center}
\includegraphics[width=8cm]{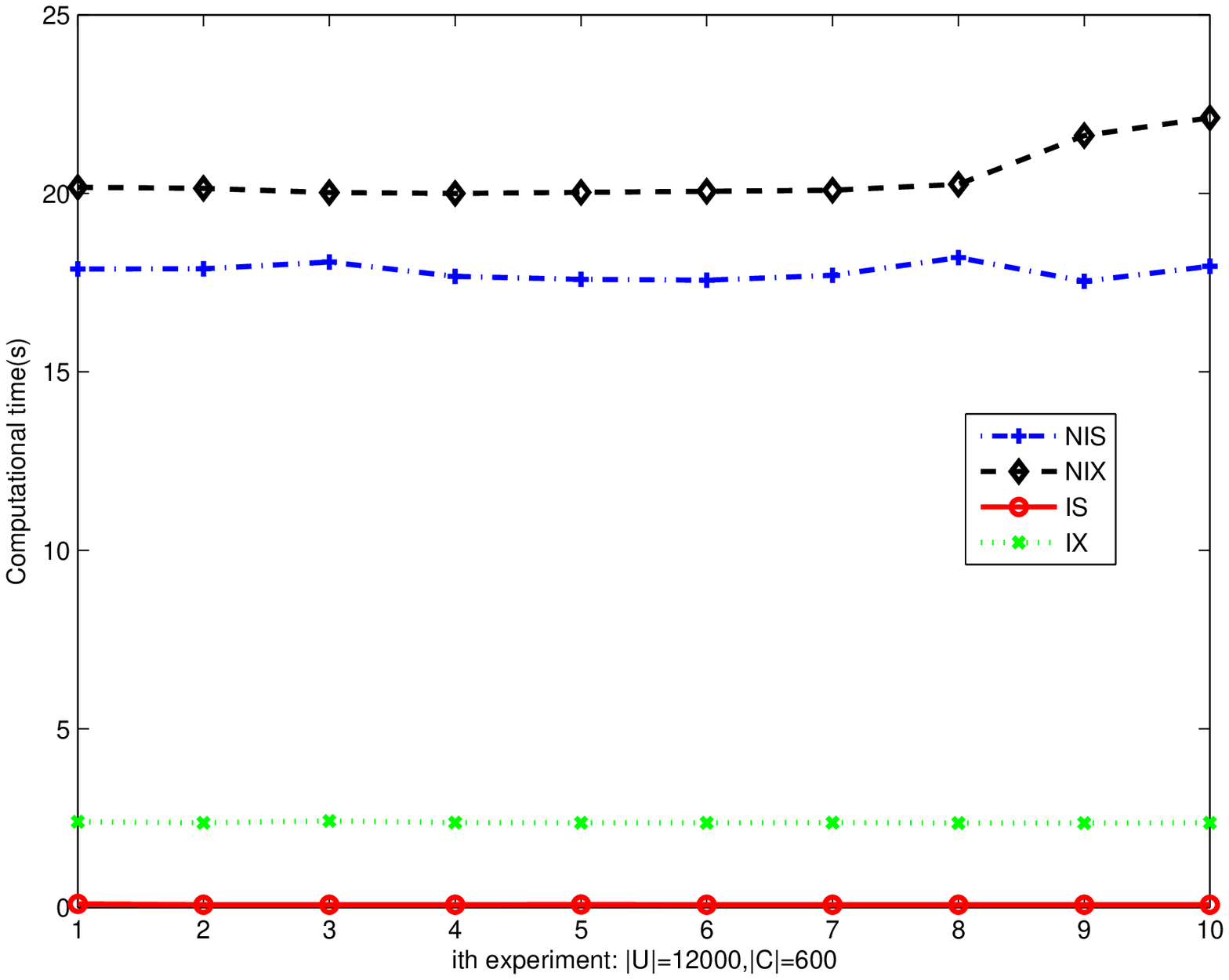}\\
\caption{Computational times using Algorithms 4.1-4.4 in
$(U_{6},\mathscr{C}_{6})$.}
\end{center}
\end{figure}

\begin{table}\renewcommand{\arraystretch}{1.5}
\caption{Computational times using Algorithms 4.1-4.4 in
$(U_{7},\mathscr{C}_{7})$. } \tabcolsep0.01in
\begin{tabular}{cccccccccccc}
\hline Algorithm&1 & 2& 3 & 4 & 5& 6& 7& 8 & 9&
10&$\overline{t}$\\\hline
 NIS&24.2936  &  24.3201  &  24.4603  &  25.2946  &  24.4922  &  24.5153  &  24.3296 &   25.0792   & 24.6210 &   24.2059   & 24.5612\\
NIX&27.9154  &  28.2049   & 28.2523  &  28.2664  &  28.7698  &  28.2559 &   28.1121 &   28.4234  &  28.6467  &  29.2779 &   28.4125\\
IS&0.1071 &    0.1014   &  0.1017  &   0.0996   &  0.1015  &   0.1018  &   0.1025  &   0.1007  &   0.1020  &   0.1009  &   0.1019\\
IX&3.4572   &  3.3194  &   3.3070  &   3.3030 &    3.2899  &   3.3109  &   3.2777  &   3.2753  &   3.2790  &   3.2758  &   3.3095\\
\hline
\end{tabular}
\end{table}

\begin{figure}
\begin{center}
\includegraphics[width=8cm]{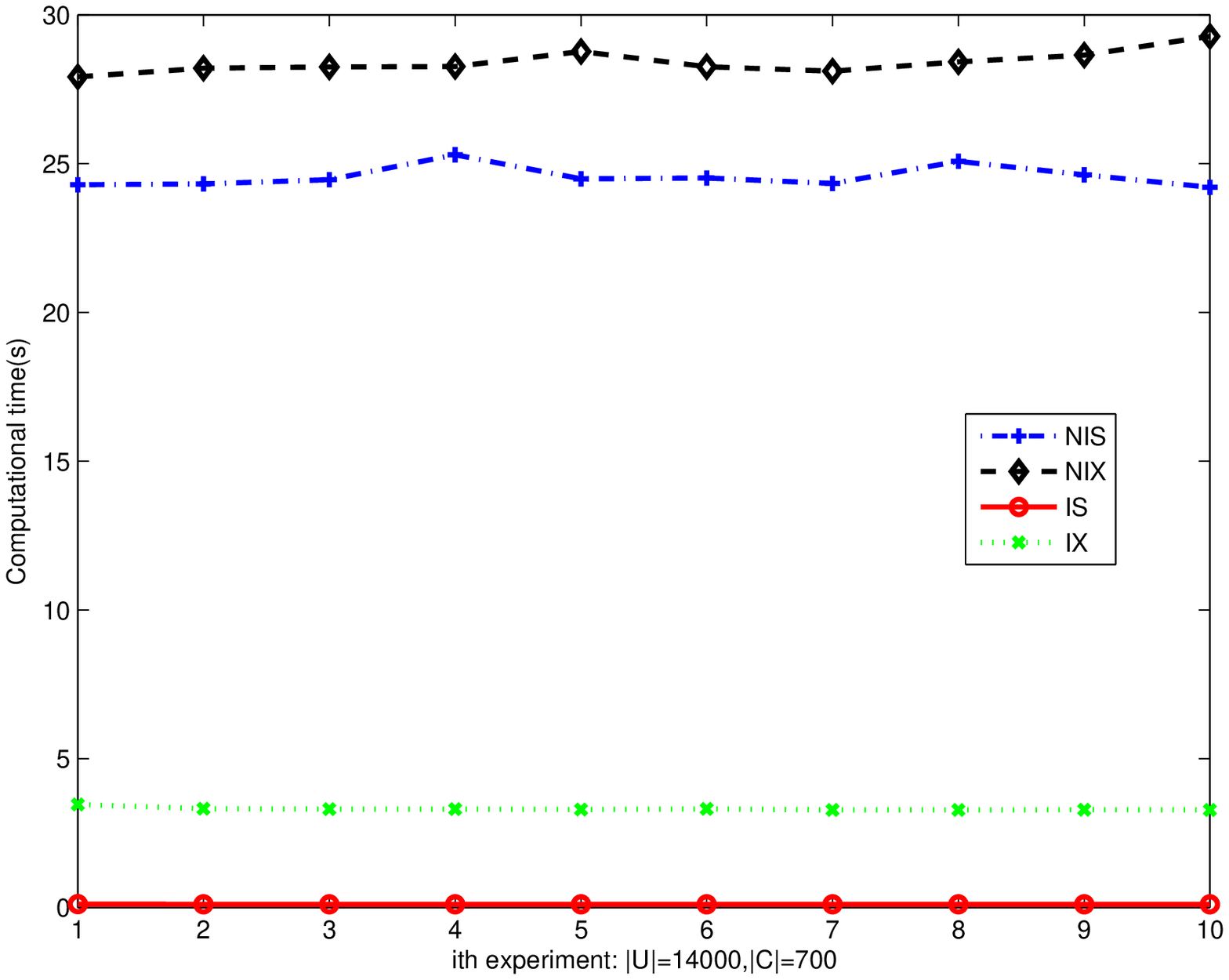}\\
\caption{Computational times using Algorithms 4.1-4.4 in
$(U_{7},\mathscr{C}_{7})$.}
\end{center}
\end{figure}
\clearpage

\begin{table}\renewcommand{\arraystretch}{1.5}
\caption{Computational times using Algorithms 4.1-4.4 in
$(U_{8},\mathscr{C}_{8})$. } \tabcolsep0.01in
\begin{tabular}{cccccccccccc}
\hline Algorithm&1 & 2& 3 & 4 & 5& 6& 7& 8 & 9&
10&$\overline{t}$\\\hline
NIS &33.2714  &  33.3024   & 33.2390  &  33.2370  &  33.3127  &  33.3602  &  33.3527  &  33.2599   & 33.4496 &   33.3485  &  33.3133\\
NIX&39.0763  &  39.0729  &  39.1256  &  39.1677 &   39.1382  &  39.5114  &  39.2732  &  38.9632   & 39.1487 &   38.8493 &   39.1327\\
IS&0.1267  &   0.1243  &   0.1293  &   0.1242  &   0.1248  &   0.1239  &   0.1259  &   0.1234   &  0.1226  &   0.1284 &    0.1254\\
IX&6.1013 &   5.3888  &   5.3412   &  5.3710    & 5.2641   &  5.3158   &  5.3229  &   5.3422   &  5.2858   &  5.4398   &  5.4173\\
\hline
\end{tabular}
\end{table}

\begin{figure}
\begin{center}
\includegraphics[width=8cm]{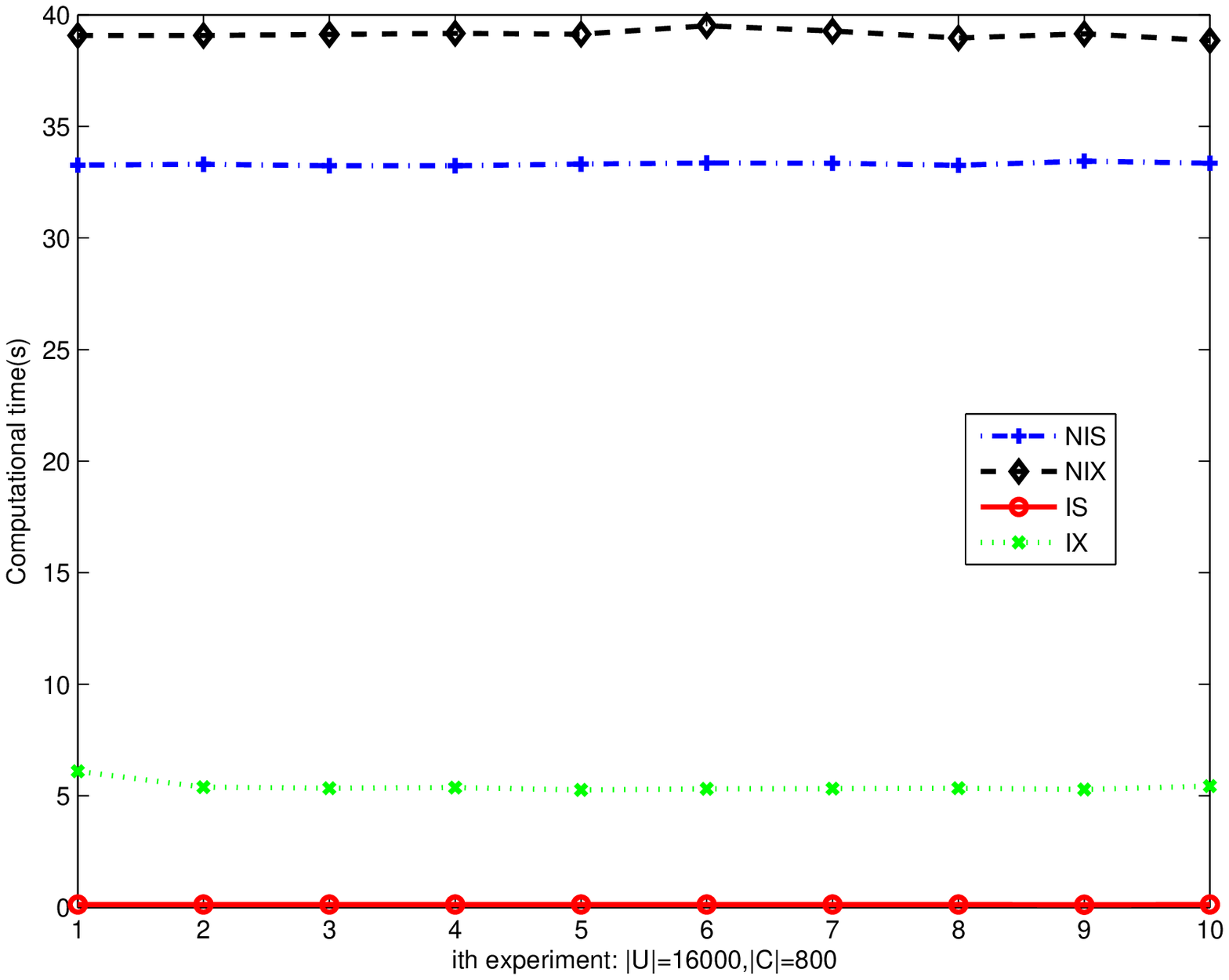}\\
\caption{Computational times using Algorithms 4.1-4.4 in
$(U_{8},\mathscr{C}_{8})$.}
\end{center}
\end{figure}

\begin{table}\renewcommand{\arraystretch}{1.5}
\caption{Computational times using Algorithms 4.1-4.4 in
$(U_{9},\mathscr{C}_{9})$. } \tabcolsep0.01in
\begin{tabular}{cccccccccccc}
\hline Algorithm&1 & 2& 3 & 4 & 5& 6& 7& 8 & 9&
10&$\overline{t}$\\\hline
NIS&44.2060  &  43.5990  &  43.2590 &   44.3375  &  43.9165 &   43.6185 &   44.3864 &   44.4667  &  44.2301 &   45.2159   & 44.1236\\
NIX&50.1711  &  50.8559   & 50.4446  &  49.7286  &  50.6871  &  50.3282  &  50.5291 &   49.5770   & 50.0544   & 50.3550  &  50.2731\\
IS&0.2048   &  0.1611  &   0.1628  &   0.1620  &   0.1607   &  0.1607  &   0.1605  &   0.1612   &  0.1615  &   0.1615 &    0.1657\\
IX&6.1794  &   5.8323  &   5.8586 &    5.7428  &   5.8902  &   5.8318  &   5.8949  &   5.7688   &  5.7606  &   5.8051   &  5.8564
\\
\hline
\end{tabular}
\end{table}

\begin{figure}
\begin{center}
\includegraphics[width=8cm]{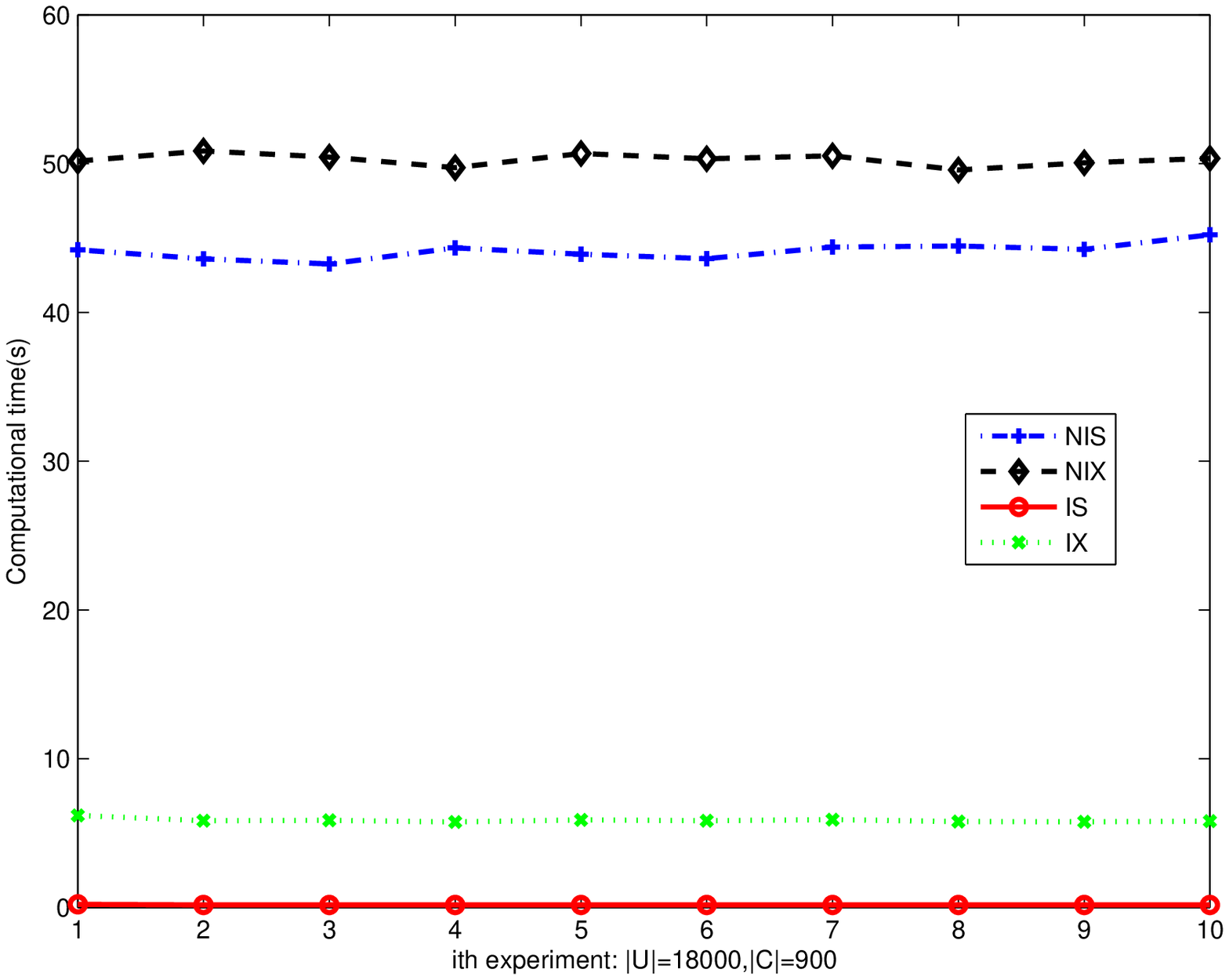}\\
\caption{Computational times using Algorithms 4.1-4.4 in
$(U_{9},\mathscr{C}_{9})$.}
\end{center}
\end{figure}

\begin{table}\renewcommand{\arraystretch}{1.5}
\caption{Computational times using Algorithms 5.1-5.8 in
$(U^{\ast}_{1},\mathscr{C}^{\ast}_{1})$, where $|U^{\ast}_{1}|=625$ and $|\mathscr{C}^{\ast}_{1}|=20$. } \tabcolsep0.01in
\begin{tabular}{cccccccccccc}
\hline Algorithm&1 & 2& 3 & 4 & 5& 6& 7& 8 & 9&
10&$\overline{t}$\\\hline
NIS&55.6793  &  55.8107 &   55.6728 &   55.9174 &   55.5917 &   58.1981  &  59.1824 &   56.0537   & 55.7757 &   55.5664 &   56.3448\\
NIX&64.8043  &  65.7104  &  65.2075 &   64.5169 &   64.7856 &   64.7118  &  65.0349   & 64.4148   & 64.7802 &   64.3155 &   64.8282\\
IS&0.2716  &   0.1941  &   0.1944  &   0.1924  &   0.1938  &   0.1956  &   0.1936  &   0.1917   &  0.1947   &  0.1948  &   0.2017\\
IX&8.3148   &  7.6287  &   7.3082  &   7.9581  &   7.2058   &  7.4084  &   7.1585  &   7.2874   &  7.1620  &   7.2413   &  7.4673
\\
\hline
\end{tabular}
\end{table}

\begin{figure}
\begin{center}
\includegraphics[width=8cm]{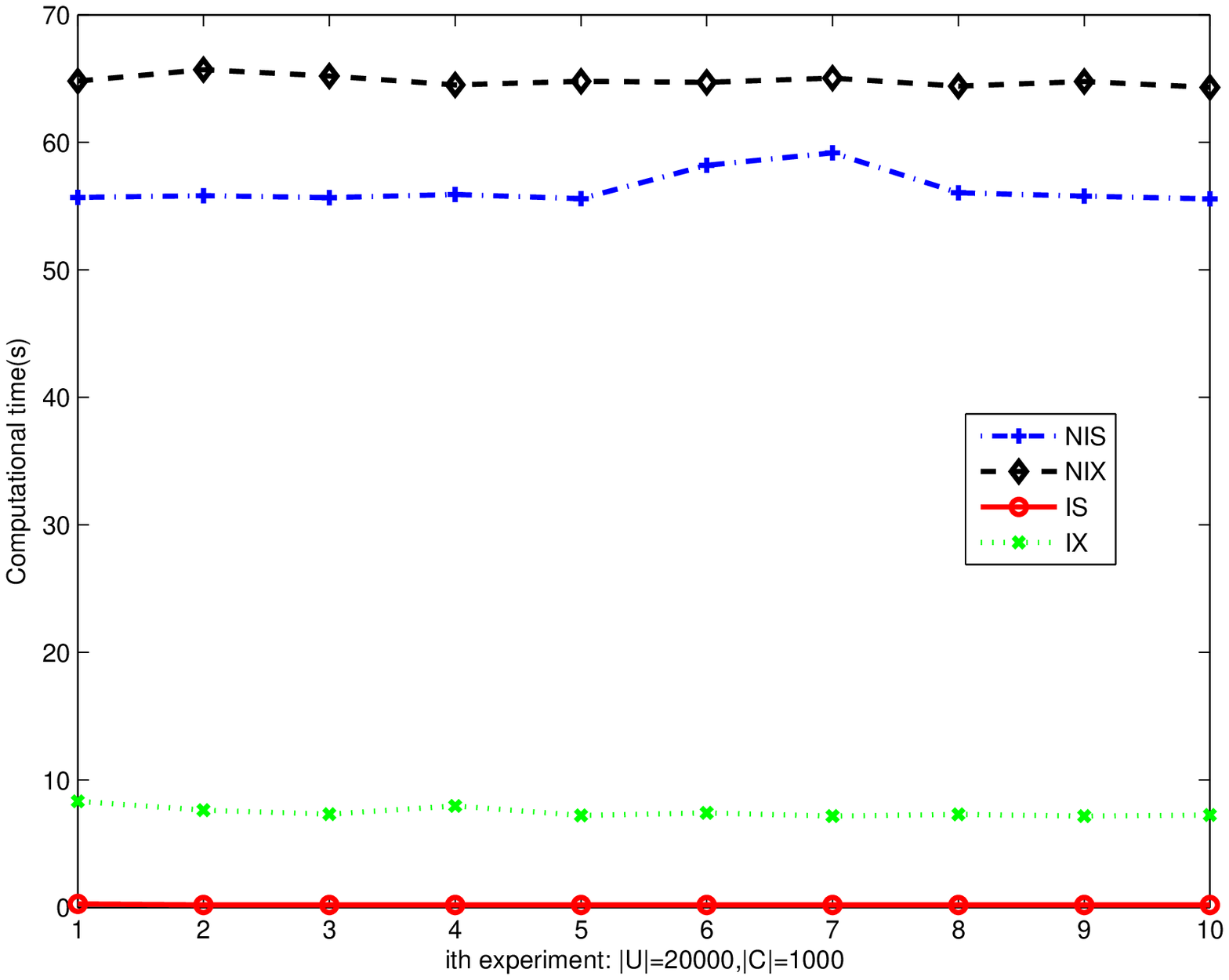}\\
\caption{Computational times using Algorithms 4.1-4.4 in
$(U_{10},\mathscr{C}_{10})$.}
\end{center}
\end{figure}

\subsubsection{The relationship between computational times and the cardinalities of object sets and coverings}

In Figure 11, the average times of the incremental
and non-incremental algorithms rise monotonically with the increase
of the cardinalities of object sets and coverings. We also see that the incremental algorithms perform always faster than the non-incremental
algorithms in all experiments, and the average times of the
incremental algorithms are much smaller than
those of the non-incremental algorithms. Moreover, the speed-up ratios of
times by using the non-incremental algorithms are higher than the
incremental algorithms with the
increasing cardinalities of object sets and coverings. Especially, we observe that there exists little influence of the cardinalities of object sets and coverings on computing the second lower and upper approximations of sets by using Algorithm 4.2.

All experimental results demonstrate that Algorithms 4.2 and 4.4 are more effective to computing the second and sixth lower and upper approximations of sets in
dynamic covering approximation spaces. In the
future, we will improve the effectiveness of Algorithms 4.2 and 4.4 and test them
on large-scale dynamic covering approximation spaces.

\begin{figure}
\begin{center}
\includegraphics[width=8cm]{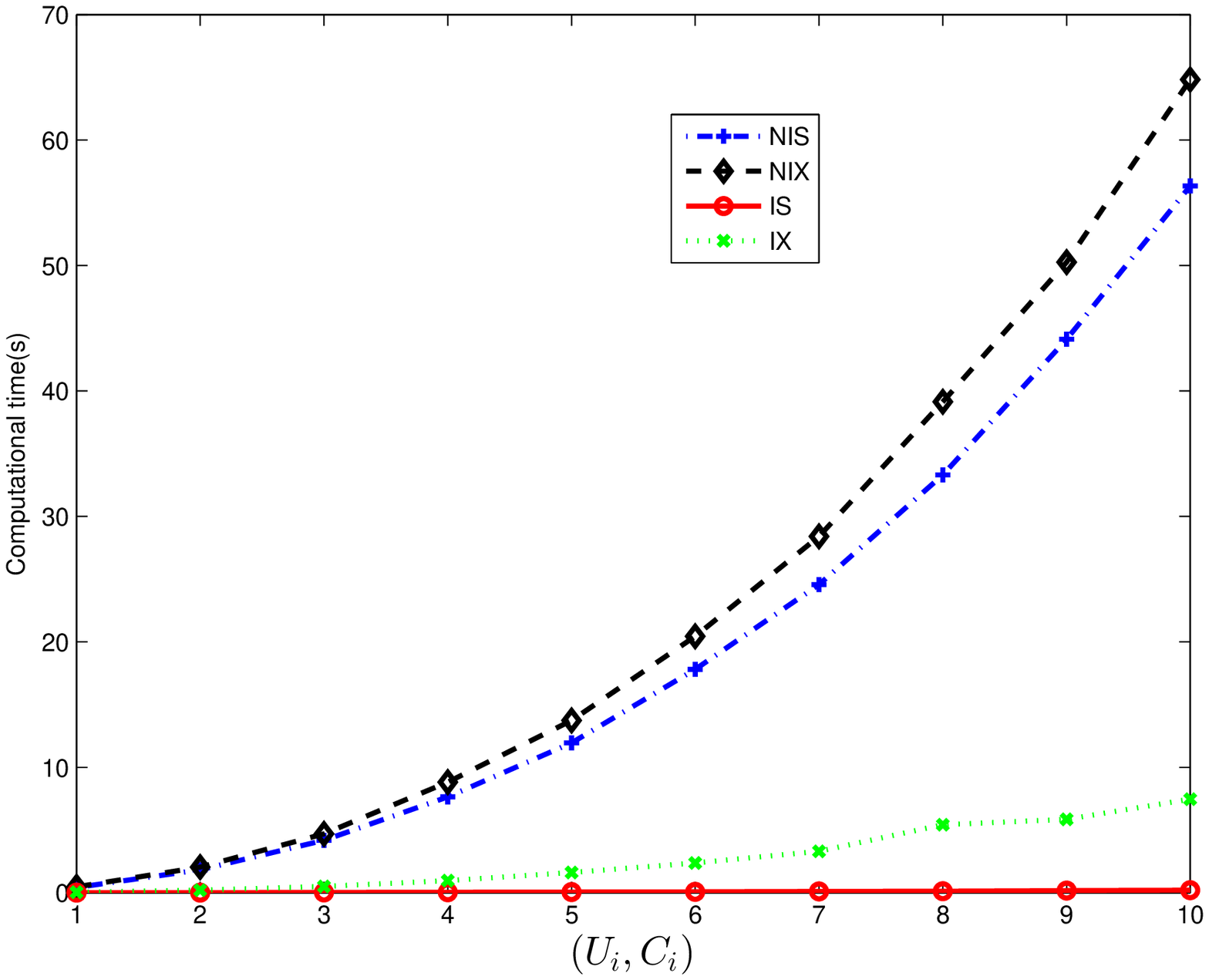}\\
\caption{Computation times using Algorithms 4.1-4.4.}
\end{center}
\end{figure}

\section{Attribute reduction of dynamic covering decision information systems}

In this section, we employ examples to illustrate that how to compute
type-1 and type-2 reducts of covering decision information systems.

\begin{example}
Let $(U,\mathscr{D}\cup U/d)$ be a covering decision information
system, where
$\mathscr{D}=\{\mathscr{C}_{1},\mathscr{C}_{2},\mathscr{C}_{3},\mathscr{C}_{4}\}$,
$\mathscr{C}_{1}=\{\{x_{1},x_{2},x_{3},x_{4}\},\{x_{5}\}\}$,
$\mathscr{C}_{2}=\{\{x_{1},x_{2}\},\{x_{3},x_{4},x_{5}\}\}$,
$\mathscr{C}_{3}=\{\{x_{1},x_{2},x_{5}\},\{x_{3},x_{4}\}\}$,
$\mathscr{C}_{4}=\{\{x_{1},x_{2}\},\{x_{3},x_{4}\},\\\{x_{5}\}\}$,
$U/d=\{\{x_{1},x_{2}\},\{x_{3},x_{4},x_{5}\}\}$. By Definitions 2.4 and 2.5,
we obtain
\begin{eqnarray*}
\Gamma(\mathscr{D})&=&\left[
\begin{array}{cccccc}
1 & 1 & 1 & 1 &1 \\
1 & 1 & 1 & 1 &1 \\
1 & 1 & 1 & 1 &1 \\
1 & 1 & 1 & 1 &1 \\
1 & 1 & 1 & 1 &1 \\
\end{array}
\right];\\
\prod(\mathscr{D})&=&\left[
\begin{array}{cccccc}
1 & 1 & 0 & 0 &0 \\
1 & 1 & 0 & 0 &0 \\
0 & 0 & 1 & 1 &0 \\
0 & 0 & 1 & 1 &0 \\
0 & 0 & 0 & 0 &1 \\
\end{array}
\right].
\end{eqnarray*}

By Definition 2.6, we have the second and sixth lower and upper
approximations of decision classes as follows:
\begin{eqnarray*} \mathcal
{X}_{SH(D_{1})}&=&\Gamma(\mathscr{D})\cdot \mathcal
{X}_{D_{1}}\\&=&\left[\begin{array}{cccccc} 1 & 1 & 1 & 1 &1
\end{array}
\right];
\\
\mathcal {X}_{SL(D_{1})}&=&\Gamma(\mathscr{D})\odot \mathcal
{X}_{D_{1}}\\&=&\left[\begin{array}{cccccc} 0 & 0 & 0 & 0 &0
\end{array}
\right];
\\
\mathcal {X}_{SH(D_{2})}&=&\Gamma(\mathscr{D})\cdot \mathcal
{X}_{D_{2}}\\&=&\left[\begin{array}{cccccc} 1 & 1 & 1 & 1 &1
\end{array}
\right];\\
\mathcal {X}_{SL(D_{2})}&=&\Gamma(\mathscr{D})\odot \mathcal
{X}_{D_{2}}\\&=&\left[\begin{array}{cccccc} 0 & 0 & 0 & 0 &0
\end{array}
\right];
\\
 \mathcal
{X}_{XH(D_{1})}&=&\Gamma(\mathscr{D})\cdot \mathcal
{X}_{D_{1}}\\&=&\left[\begin{array}{cccccc} 1 & 1 & 0 & 0 & 0
\end{array}
\right];
\\
\mathcal {X}_{XL(D_{1})}&=&\Gamma(\mathscr{D})\odot \mathcal
{X}_{D_{1}}\\&=&\left[\begin{array}{cccccc} 1 & 1 & 0 & 0 & 0
\end{array}
\right];
\\
\mathcal {X}_{XH(D_{2})}&=&\Gamma(\mathscr{D})\cdot \mathcal
{X}_{D_{2}}\\&=&\left[\begin{array}{cccccc} 0 & 0 & 1 & 1 & 1
\end{array}
\right];\\
\mathcal {X}_{XL(D_{2})}&=&\Gamma(\mathscr{D})\odot \mathcal
{X}_{D_{2}}\\&=&\left[\begin{array}{cccccc} 0 & 0 & 1 & 1 & 1
\end{array}
\right].
\end{eqnarray*}

To construct type-1 and type-2 reducts, we have that
\begin{eqnarray*}
\Gamma(\mathscr{D}/\mathscr{C}_{4})\cdot \mathcal
{X}_{D_{1}}&=&\mathcal
{X}_{SH(D_{1})};\\
\Gamma(\mathscr{D}/\mathscr{C}_{4})\odot \mathcal
{X}_{D_{1}}&=&\mathcal
{X}_{SL(D_{1})};\\\Gamma(\mathscr{D}/\mathscr{C}_{4})\cdot \mathcal
{X}_{D_{2}}&=&\mathcal {X}_{SH(D_{2})};\\
\Gamma(\mathscr{D}/\mathscr{C}_{4})\odot \mathcal
{X}_{D_{2}}&=&\mathcal {X}_{SL(D_{2})};\\
\prod(\mathscr{D}/\mathscr{C}_{4})\cdot \mathcal
{X}_{D_{1}}&=&\mathcal
{X}_{XH(D_{1})};\\
\prod(\mathscr{D}/\mathscr{C}_{4})\odot \mathcal
{X}_{D_{1}}&=&\mathcal {X}_{XL(D_{1})};
\\
\prod(\mathscr{D}/\mathscr{C}_{4})\cdot \mathcal
{X}_{D_{2}}&=&\mathcal
{X}_{XH(D_{2})};\\
\prod(\mathscr{D}/\mathscr{C}_{4})\odot \mathcal
{X}_{D_{2}}&=&\mathcal {X}_{XL(D_{2})};
\end{eqnarray*}

To perform the above process continuously, we have that
$\{\mathscr{C}_{1},\mathscr{C}_{3}\}$ is type-1 and type-2 reducts
of $(U,\mathscr{D}\cup U/d)$.
\end{example}

We employ an example to illustrate that how to construct type-1 and
type-2 reducts of dynamic covering decision information systems as
follows.

\begin{example}(Continuation of Example 6.1)
Let $(U,\mathscr{D}^{\ast}\cup U/d)$ be a covering decision
information system, where
$\mathscr{D}^{\ast}=\{\mathscr{C}^{\ast}_{1},\mathscr{C}^{\ast}_{2},\mathscr{C}^{\ast}_{3},\mathscr{C}^{\ast}_{4}\}$,
$\mathscr{C}^{\ast}_{1}=\{\{x_{1},x_{2},x_{3},x_{4}\},\{x_{5}\}\}$,
$\mathscr{C}^{\ast}_{2}=\{\{x_{1},x_{2}\},\{x_{3},x_{4},x_{5}\}\}$,
$\mathscr{C}^{\ast}_{3}=\{\{x_{1},x_{2},x_{3},x_{5}\},\{x_{4}\}\}$,
$\mathscr{C}^{\ast}_{4}=\{\{x_{1},x_{2}\},\{x_{3},x_{4}\},\{x_{5}\}\}$,
$U/d=\{\{x_{1},x_{2}\},\{x_{3},x_{4},x_{5}\}\}$. By Theorems 3.2 and
3.4, we obtain
\begin{eqnarray*}
\Gamma(\mathscr{D}^{\ast})&=&\left[
\begin{array}{cccccc}
1 & 1 & 1 & 1 &1 \\
1 & 1 & 1 & 1 &1 \\
1 & 1 & 1 & 1 &1 \\
1 & 1 & 1 & 1 &1 \\
1 & 1 & 1 & 1 &1 \\
\end{array}
\right];\\
\prod(\mathscr{D}^{\ast})&=&\left[
\begin{array}{cccccc}
1 & 1 & 0 & 0 &0 \\
1 & 1 & 0 & 0 &0 \\
0 & 0 & 1 & 0 &0 \\
0 & 0 & 0 & 1 &0 \\
0 & 0 & 0 & 0 &1 \\
\end{array}
\right].
\end{eqnarray*}

By Definition 2.6, we have the second and sixth lower and upper
approximations of decision classes as follows:
\begin{eqnarray*} \mathcal
{X}_{SH(D_{1})}&=&\Gamma(\mathscr{D}^{\ast})\cdot \mathcal
{X}_{D_{1}}\\&=&\left[\begin{array}{cccccc} 1 & 1 & 1 & 1 &1
\end{array}
\right];
\\
\mathcal {X}_{SL(D_{1})}&=&\Gamma(\mathscr{D}^{\ast})\odot \mathcal
{X}_{D_{1}}\\&=&\left[\begin{array}{cccccc} 0 & 0 & 0 & 0 &0
\end{array}
\right];
\\
\mathcal {X}_{SH(D_{2})}&=&\Gamma(\mathscr{D}^{\ast})\cdot \mathcal
{X}_{D_{2}}\\&=&\left[\begin{array}{cccccc} 1 & 1 & 1 & 1 &1
\end{array}
\right];\\
\mathcal {X}_{SL(D_{2})}&=&\Gamma(\mathscr{D}^{\ast})\odot \mathcal
{X}_{D_{2}}\\&=&\left[\begin{array}{cccccc} 0 & 0 & 0 & 0 &0
\end{array}
\right];\\
 \mathcal
{X}_{XH(D_{1})}&=&\Gamma(\mathscr{D}^{\ast})\cdot \mathcal
{X}_{D_{1}}\\&=&\left[\begin{array}{cccccc} 1 & 1 & 0 & 0 & 0
\end{array}
\right];
\\
\mathcal {X}_{XL(D_{1})}&=&\Gamma(\mathscr{D}^{\ast})\odot \mathcal
{X}_{D_{1}}\\&=&\left[\begin{array}{cccccc} 1 & 1 & 0 & 0 & 0
\end{array}
\right];
\\
\mathcal {X}_{XH(D_{2})}&=&\Gamma(\mathscr{D}^{\ast})\cdot \mathcal
{X}_{D_{2}}\\&=&\left[\begin{array}{cccccc} 0 & 0 & 1 & 1 & 1
\end{array}
\right];\\
\mathcal {X}_{XL(D_{2})}&=&\Gamma(\mathscr{D}^{\ast})\odot \mathcal
{X}_{D_{2}}\\&=&\left[\begin{array}{cccccc} 0 & 0 & 1 & 1 & 1
\end{array}
\right].
\end{eqnarray*}

To construct type-1 and type-2 reducts, we have that
\begin{eqnarray*}
\Gamma(\mathscr{D}^{\ast}/\mathscr{C}^{\ast}_{4})\cdot \mathcal
{X}_{D_{1}}&=&\mathcal
{X}_{SH(D_{1})};\\
\Gamma(\mathscr{D}^{\ast}/\mathscr{C}^{\ast}_{4})\odot \mathcal
{X}_{D_{1}}&=&\mathcal
{X}_{SL(D_{1})};\\\Gamma(\mathscr{D}^{\ast}/\mathscr{C}^{\ast}_{4})\cdot
\mathcal
{X}_{D_{2}}&=&\mathcal {X}_{SH(D_{2})};\\
\Gamma(\mathscr{D}^{\ast}/\mathscr{C}^{\ast}_{4})\odot \mathcal
{X}_{D_{2}}&=&\mathcal {X}_{SL(D_{2})};\\
\prod(\mathscr{D}^{\ast}/\mathscr{C}^{\ast}_{4})\cdot \mathcal
{X}_{D_{1}}&=&\mathcal
{X}_{XH(D_{1})};\\
\prod(\mathscr{D}^{\ast}/\mathscr{C}^{\ast}_{4})\odot \mathcal
{X}_{D_{1}}&=&\mathcal {X}_{XL(D_{1})};
\\
\prod(\mathscr{D}^{\ast}/\mathscr{C}^{\ast}_{4})\cdot \mathcal
{X}_{D_{2}}&=&\mathcal
{X}_{XH(D_{2})};\\
\prod(\mathscr{D}^{\ast}/\mathscr{C}^{\ast}_{4})\odot \mathcal
{X}_{D_{2}}&=&\mathcal {X}_{XL(D_{2})};
\end{eqnarray*}

To perform the above process continuously, we have that
$\{\mathscr{C}^{\ast}_{1},\mathscr{C}^{\ast}_{3}\}$ is a type-1
reduct of $(U,\mathscr{D}^{\ast}\cup U/d)$, and
$\{\mathscr{C}^{\ast}_{1},\mathscr{C}^{\ast}_{2},\mathscr{C}^{\ast}_{3}\}$
is a type-2 reduct of $(U,\mathscr{D}^{\ast}\cup U/d)$.
\end{example}

\section{Conclusions}

Knowledge reduction of covering information systems have attracted more attention of researchers.
In this paper, we have introduced incremental
approaches to computing the characteristic matrices of dynamic coverings when revising attribute values. We have presented the non-incremental and incremental algorithms for computing the second and sixth lower and upper approximations of sets and
compared the computational complexities of the non-incremental algorithms with those of incremental algorithms. We have tested the incremental algorithms on dynamic covering approximation spaces.
Experimental results have been employed to illustrate that the
incremental approaches are effective to compute approximations of sets in dynamic covering approximation spaces. We have demonstrated that how to conduct knowledge reduction of dynamic covering information systems with the incremental approaches.

In practical situations, there exist many types of dynamic covering information systems and dynamic covering approximation spaces. In the future,  we will introduce more effective approaches to
constructing the characteristic matrices of these types of dynamic coverings and perform knowledge reduction of these types of dynamic covering information systems.

\section*{ Acknowledgments}

We would like to thank the anonymous reviewers very much for their
professional comments and valuable suggestions. This work is
supported by the National Natural Science Foundation of China (NO.
11201490,11371130,11401052,11401195), the Scientific Research Fund of Hunan
Provincial Education Department(No.14C0049).

\end{document}